\documentclass[english,aps,pra,twocolumn,superscriptaddress]{revtex4-2}
\usepackage[utf8]{inputenc}
\usepackage{color}
\usepackage{babel}
\usepackage{mathtools}
\usepackage{bm}
\usepackage{amsmath}
\usepackage{amssymb}
\usepackage{graphicx}
\usepackage[colorlinks=true,citecolor=blue]{hyperref}

\makeatletter
\usepackage{xcolor}
\usepackage{babel}
\usepackage{bm}
\usepackage{physics}
\usepackage{empheq}
\usepackage{times}

\makeatother

\begin{document}
\title{Optimal quantum state tomography with local informationally complete measurements}

\author{Casey Jameson}
\email{cwjameson@mines.edu}
\affiliation{Department of Physics, Colorado School of Mines, Golden, Colorado 80401, USA}

\author{Zhen Qin}
\affiliation{Department of Computer Science and Engineering, The Ohio State University, Columbus, OH 43210, USA}

\author{Alireza Goldar}
\affiliation{Department of Electrical Engineering, Colorado School of Mines, Golden, Colorado 80401, USA}

\author{Michael B. Wakin}
\affiliation{Department of Electrical Engineering, Colorado School of Mines, Golden, Colorado 80401, USA}

\author{Zhihui Zhu}
\affiliation{Department of Computer Science and Engineering, The Ohio State University, Columbus, OH 43210, USA}

\author{Zhexuan Gong}
\email{gong@mines.edu}
\affiliation{Department of Physics, Colorado School of Mines, Golden, Colorado 80401, USA}
\affiliation{National Institute of Standard and Technology, Boulder, Colorado 80305, USA}

\begin{abstract}
Quantum state tomography (QST) remains the gold standard for benchmarking and verification of quantum computers and quantum simulators. The current scale of experimental quantum devices already makes direct quantum state tomography intractable due to the exponentially large number of parameters in a generic quantum many-body state. However, most physical quantum states are structured and can often be represented by a much smaller number of parameters, making efficient QST potentially possible. A prominent example is a matrix product state (MPS) or a matrix product density operator (MPDO) with a small matrix dimension, which is believed to represent most physical states generated by one-dimensional (1D) quantum devices. We study whether a general MPS/MPDO state can be recovered with bounded errors using only a number of state copies polynomial in the number of qubits, which is necessary for efficient QST. To make this question practically interesting, we assume only local measurements of qubits directly on the target state. By using a local symmetric informationally complete positive operator-valued measurement (SIC-POVM) that requires only a single measurement setting, we provide a positive answer to the above question for a variety of common many-body quantum states, including typical short-range entangled states, random MPS/MPDO states, and thermal states of one-dimensional Hamiltonians. In addition, we also provide an affirmative no answer for certain long-range entangled states such as a family of generalized GHZ states, but with the exception of target states that are known to have real-valued wavefunctions. Our answers are supported by a near-perfect agreement between an efficient calculation of the Cramer-Rao bound that rigorously bounds the sample complexity and numerical optimization results using a machine learning assisted maximal likelihood estimation (MLE) algorithm. This agreement also leads to an optimal QST protocol using local SIC-POVM that can be practically implemented on current quantum hardware and is highly efficient for most 1D physical states. Our results also indicate that long-range entangled quantum states are in general not efficiently recoverable even if they are efficiently representable.
\end{abstract}
\maketitle

\section{Introduction}

With quantum computers rapidly scaling in size from the small-scale devices composed of only a handful of qubits to the current intermediate-scale devices composed of tens or even hundreds of qubits \cite{Preskill_2018,IonQComputer2021,IBMComputer2023}, the challenge of validating the performance of those devices scales at a similarly rapid pace. While a number of methods to verify the quantum state generated by intermediate-scale quantum devices have been proposed \cite{Aaronson2020,Huang2020,Harney2020,Zhu2022,Bu2024}, the gold standard remains quantum state tomography (QST) \cite{Cramer2010,Flammia2012,Baumgratz2013,BaumgratzMLE2013,Franca2020}, as it provides complete information about the experimental state. However, for a generic and possibly mixed $N$-qubit state, the number of state copies required to recover the state scales as $\mathcal{O}(2^{N})$ even with the optimal measurement setting \cite{Haah2017,Guta2020}. Therefore, QST for generic quantum states is not scalable and only affordable on small-scale quantum devices.

In recent years, significant effort has been made toward the development of scalable QST methods by exploiting the fact that most physical states are structured in certain ways, allowing efficient classical representations through models such as matrix product states (MPS) \cite{perez-garcia_matrix_2007}, matrix product operators (MPO) \cite{verstraete_matrix_2004}, tensor network states \cite{silvi_tensor_2019}, and quantum neural networks \cite{Carleo2017,Schmale2022,Innan2024}. However, efficient representation of a state is in general not sufficient for efficient QST, which requires the number of state copies, the number of elementary quantum gates needed for measurement, and the classical post-processing time to be all polynomial in $N$. While finding sufficient conditions for efficient QST is still a major challenge in quantum information theory, a practically more important challenge is to find efficient QST protocols that work for most physical states obtained on a given experimental quantum hardware.

In this paper, we attempt to address the above challenges with a focus on quantum states efficiently representable by MPS/MPO. These states incorporate most of the physical states generated by one-dimension quantum computers or quantum simulators, including those with noise and errors \cite{noh_efficient_2020}. An efficient representation here implies that the dimension of the matrices (often known as the bond dimension) in the MPS/MPO is either a constant or grows slowly with $N$, such that only a number of parameters polynomial in $N$ is used to represent the state. Previous studies \cite{Baumgratz2013,Cramer2010,BaumgratzMLE2013} show that efficient QST is possible if the MPS/MPO is locally reducible, meaning that the full system's state can be reconstructed with bounded errors using only reduced states on subsystems made of a small, $N$-independent number of qubits. Here efficient QST also assumes that the target state is reconstructed by an MPS/MPO with a small bond dimension. While the full system's wavefunction or density matrix cannot be reconstructed efficiently, most physically interesting properties, including non-local ones such as the entanglement entropy or topological order parameters, can be efficiently computed using the MPS/MPO representation of the state. In this regard, MPS/MPO based QST is more powerful than shadow tomography or randomized measurement toolbox \cite{huang_predicting_2020,elben_randomized_2022} that can usually only predict few-body correlations efficiently.

However, there are many physically interesting quantum states that are not locally reducible, such as certain long-range entangled states \cite{chen_local_2010} and topologically ordered states \cite{bravyi_lieb-robinson_2006,tran_hierarchy_2020}. More importantly, mixed states, ubiquitous experimentally, are in general not locally reducible since those reduced states often do not uniquely determine the full system's state \cite{Cramer2010}. A complicated mathematical condition is needed to determine if a mixed state represented by an MPO is locally reducible \cite{baumgratz_scalable_2013}. Therefore, it remains unclear if efficient QST is possible for MPS/MPO states that are not locally reducible. In a recent work \cite{Qin2024}, we show that if one can measure all qubits in a global, Haar random basis, then any MPS/MPO with a finite bond dimension can be recovered with bounded errors using a number of state copies polynomial in $N$. This, however, does not imply efficient QST since the global Haar random measurement cannot be implemented efficiently. Therefore, here we restrict to only local measurement, meaning that each qubit is measured independently and individually by some POVM. Such measurement can be readily implemented in most current quantum hardware.

The first main contribution of this work is the development of an efficient method to identify the minimum number of state copies $M$ required to perform QST on any given target state with an efficient MPS/MPO representation for a local, informationally complete POVM. This method is based on the Cramer-Rao bound in estimation theory \cite{Pintelon1996} that lower bounds the statistical error of estimating parameters in a given probability distribution. While the Cramer-Rao bound has been widely used in evaluating the performance of various QST protocols \cite{GS2000,Usami2003,kosut2004,Wang2020}, it is usually only evaluated for very small quantum systems. We show that by utilizing the structure of MPS/MPO, the Cramer-Rao bound can be computed efficiently via a novel autoregressive Monte Carlo sampling method for a large quantum many-body system. With this method, we show that typical short-range entangled states \cite{Chen2007} and random MPS/MPO states can indeed be recovered efficiently, using a number of state copies proportional to the number of independent parameters in the target state. On the other hand, we reveal that there are certain long-range entangled states, e.g. a family of generalized GHZ states, that requires an exponentially large (in $N$) number of state copies to be reconstructed with a finite error. But surprisingly, if the target state is known to be represented by a real-valued MPS/MPO, this exponential scaling vanishes, despite the fact that the target state is still not locally reducible. Therefore, we argue that local reducibility does not by itself determine whether efficient QST via MPS/MPO is possible.

The second main contribution of our work is the development of a new QST protocol that combines the MPS/MPO model, local SIC-POVM, and a machine learning assisted MLE algorithm. We show that this protocol is optimal as it saturates the Cramer-Rao bound for a variety of physical target states within statistical errors.  Theoretically, this optimality comes from the fact that MLE saturates the Cramer-Rao bound asymptotically in the large sample size limit \cite{Braunstein1992,Sugiyama2011,Christandl2012,Qi2013,Scholten2018}. There are also several practical advantages of our protocol: First, the use of local SIC-POVM allows us to measure the quantum state in a single measurement setting/basis for any system size \cite{Renes2004,Scott2006,Słomczynski2015,Yoshida2022}, in contrast to the more common local Pauli measurement that requires $3^{N}$ different measurement settings (i.e. measuring $\sigma_{i}^{x,y,z}$ for each qubit $i$). This removes the need for individual addressing of the qubits (individual readout of the qubits is still required), which can be significantly beneficial for certain quantum simulation platforms. For generic states, QST using SIC-POVM also requires less number of state copies than that based on the Pauli measurement for the same accuracy \cite{Guta2020}. Moreover, for efficient QST, one cannot implement all $3^{N}$ measurement settings and the use of local Pauli measurement thus does not guarantee information completeness. While one can randomly pick the measurement basis \cite{Wang2020} in hope of gaining information completeness statistically, this approach does not perform as well as our protocol and makes it challenging to evaluate the Cramer-Rao bound due to the randomness in the measurement settings. Experimentally, local SIC-POVM can be performed by mapping each qubit to a ququart using either ancilla qubits or extra states outside the qubit subspace, which has been recently demonstrated using trapped ions in a scalable way \cite{Stricker2022}.

Second, a single measurement setting also allows us to evaluate physically interesting quantities in real time as experimental measurement samples are collected. Since our protocol does not rely on local reductions, one can estimate global quantities such as entanglement entropy or topological order parameters. The experimental state preparation and measurement cycles can be stopped once a desired convergence is reached \cite{Stricker2022}. Finally, by using state-of-the-art machine-learning methods, our MLE algorithm has an affordable classical runtime with efficient calculations of the gradients. It also works well for both pure and mixed target states, in contrast to many MPS based QST algorithms that cannot be easily adapted to mixed states \cite{Wang2020,Cramer2010}. We have also benchmarked our algorithm using synthetic measurement data on a variety of common 1D physical states including typical short-range entangled states, random MPS/MPO states, and thermal states of 1D Hamiltonians. In most cases, the number of state copies needed to achieve high-fidelity QST is proportional to the number of parameters in the MPS/MPO representation of the target state, which ensures efficient QST.

The remainder of this paper is organized as follows: In Section \ref{sec:Model}, we introduce the measurement setup (local SIC-POVM) used in this work and the model (MPS/MPO) used to reconstruct the target state. In Section \ref{sec:Cramer-Rao-bounds}, we introduce the Cramer-Rao bound and show how to calculate it efficiently to lower bound the number of state copies needed to reconstruct a given target state. We compute the Cramer-Rao bound for various many-body entangled states with efficient MPS representations explicitly, revealing that efficient QST is possible for some target states but not the others. In Section \ref{ExpResultsSection}, we introduce our MLE algorithm and benchmark it for a range of physical states in one dimension, showing that the fidelity of QST obtained using synthetic measurement samples closely matches that predicted by the Cramer-Rao bound. The paper concludes with an outlook on the remaining open questions.

\section{Model and Measurement Setting}\label{sec:Model}

We start by introducing the measurement setting used throughout this paper, which is a local SIC-POVM on each qubit. SIC-POVM is a particular type of generalized quantum measurement that is highly symmetric, information complete, and having the minimal number of outcomes compatible with informational completeness. For a single qubit, a commonly used example of SIC-POVM is defined by four positive operators $M_{i}=|\phi_{i}\rangle\langle\phi_{i}|$ ($i=1,2,3,4$) with
\begin{align}
|\phi_{1}\rangle & =\frac{1}{\sqrt{2}}|0\rangle\label{eq:SIC-POVM}\\
|\phi_{2}\rangle & =\frac{1}{\sqrt{6}}|0\rangle+\frac{1}{\sqrt{3}}|1\rangle\nonumber \\
|\phi_{3}\rangle & =\frac{1}{\sqrt{6}}|0\rangle+\frac{1}{\sqrt{3}}e^{i\frac{2\pi}{3}}|1\rangle\nonumber \\
|\phi_{4}\rangle & =\frac{1}{\sqrt{6}}|0\rangle+\frac{1}{\sqrt{3}}e^{-i\frac{2\pi}{3}}|1\rangle\nonumber 
\end{align}
It is easy to check that $\sum_{i}M_{i}=\hat{1}$ and that $\{M_{i}\}$ spans the space of any single-qubit density operator. Therefore, $\{M_{i}\}$ form an informationally complete (IC) POVM. The symmetry of this IC-POVM manifests in $\text{Tr}(M_{i}M_{j})=\frac{d\delta_{ij}+1}{d^{2}(d+1)}$ for any $i$ and $j$, where $d=2$ for qubit.

Experimentally, the above SIC-POVM can be implemented by mapping the state of a qubit (with basis states $|0\rangle$ and $|1\rangle$) to a state of a ququart (with basis states $|0\rangle$, $|1\rangle$, $|2\rangle$, and $|3\rangle$) and performing a standard projective measurement for the ququart. The extra basis states ($|2\rangle$ and $|3\rangle$) can either come from additional states in the physical system carrying the qubit \cite{Stricker2022}, or from an auxiliary qubit \cite{Chen2007,Garcia2021}. The mapping from qubit to ququart can be done by applying the following unitary transformation on the ququart, starting from any initial state in the qubit subspace
\begin{align}
U_{\text{SIC}} & =\left(\begin{array}{cccc}
\frac{1}{\sqrt{2}} & 0 & 0 & \frac{1}{\sqrt{2}}\\
\frac{1}{\sqrt{6}} & \frac{1}{\sqrt{3}} & \frac{1}{\sqrt{3}} & -\frac{1}{\sqrt{6}}\\
\frac{1}{\sqrt{6}} & \frac{e^{-i\frac{2\pi}{3}}}{\sqrt{3}} & \frac{e^{i\frac{2\pi}{3}}}{\sqrt{3}} & -\frac{1}{\sqrt{6}}\\
\frac{1}{\sqrt{6}} & \frac{e^{i\frac{2\pi}{3}}}{\sqrt{3}} & \frac{e^{-i\frac{2\pi}{3}}}{\sqrt{3}} & -\frac{1}{\sqrt{6}}
\end{array}\right).\label{eq:USIC}
\end{align}
This unitary can be achieved by a combination of pulses driving the $|0\rangle-|1\rangle$, $|0\rangle-|2\rangle$, and $|0\rangle-|3\rangle$ transitions respectively \cite{Stricker2022}. Next, we measure the ququart projectively with its state collapsing to either $|0\rangle$, $|1\rangle$, $|2\rangle$, or $|3\rangle$. Denote the initial state of the qubit by $\rho=\begin{pmatrix}\rho_{00} & \rho_{01}\\ \rho_{10} & \rho_{11} \end{pmatrix}$, it's not hard to see that the probabilities for the ququart measurement are given by
\begin{align}
P(|0\rangle) & =\frac{1}{2}\rho_{00}\label{eq:PSIC} \\
P(|1\rangle) & =\frac{1}{6}\left(\rho_{00}+2\rho_{11}+\sqrt{2}\rho_{01}+\sqrt{2}\rho_{10}\right)\nonumber \\
P(|2\rangle) & =\frac{1}{6}\left(\rho_{00}+2\rho_{11}+\sqrt{2}e^{i\frac{2\pi}{3}}\rho_{01}+\sqrt{2}e^{-i\frac{2\pi}{3}}\rho_{10}\right)\nonumber \\
P(|3\rangle) & =\frac{1}{6}\left(\rho_{00}+2\rho_{11}+\sqrt{2}e^{-i\frac{2\pi}{3}}\rho_{01}+\sqrt{2}e^{i\frac{2\pi}{3}}\rho_{10}\right) \nonumber
\end{align}

The qubit's state $\rho$ can thus be reconstructed from these four probabilities directly by solving the above set of linear equations. Since most quantum hardware platforms have single-shot readout of the qubit's state, a single-shot ququart readout can be experimentally achieved by three sequential qubit readouts with the help of $\pi$-pulses between the states $|2\rangle$ and $|3\rangle$ and the state $|0\rangle$ or $|1\rangle$ \cite{Stricker2022}.

For multiple qubits, we perform the above SIC-POVM for each qubit independently, either in parallel or sequentially (since the measurements on different qubits mutually commute). The above unitary $U_{\text{SIC}}$ for mapping each qubit to a ququart can also be implemented in parallel without individual addressing of the qubits. Therefore, our measurement setup requires only a constant circuit depth and is scalable.

Next, we introduce the class of quantum states we are targeting for our QST protocol. We consider a general target state represented by the density operator $\rho=\sum_{\bm{s},\bm{s}^{\prime}}\rho_{\bm{s},\bm{s}^{\prime}}|\bm{s}\rangle\langle\bm{s}^{\prime}|$ expanded in the standard qubit basis $|\bm{s}\rangle\equiv|s_{1}s_{2}\cdots s_{N}\rangle$ with $s_{i}=0,1$ for $i=1,2,\cdots,N$. The class of states we focus on in this work takes the form of an MPO:
\begin{equation}
\rho_{\bm{s},\bm{s}^{\prime}}=\text{Tr}(A_{1}^{s_{1}s_{1}^{\prime}}A_{2}^{s_{2}s_{2}^{\prime}}\cdots A_{N}^{s_{N}s_{N}^{\prime}})\label{eq:MPO}
\end{equation}
where each $A_{i}^{s_{i}s_{i}^{\prime}}$ is a square matrix of dimension $\eta$ (the assumption of square matrix here is more for convenience and can be relaxed). For the state to have an efficient MPO representation, we require $\eta$ to be increasing at most polynomially in $N$. For simplicity, we assume $\eta$ is a constant for the rest of the paper, as recent work suggests that states generated by a one-dimensional noisy quantum circuit can indeed be well approximated by MPOs with $\eta$ independently of $N$ \cite{Noh2020}. Our results below can be straightforwardly generalized to the cases where $\eta$ scales with $N$.

To make sure the MPO in the above Eq.\,\eqref{eq:MPO} represents a physical density operator that is always Hermitian and semi-positive definite, we further assume that each matrix $A_{i}^{s_{i}s_{i}^{\prime}}$ can be expressed as
\begin{equation}
A_{i}^{s_{i}s_{i}^{\prime}}=\sum_{k=1}^{\kappa}C_{i,k}^{s_{i}}\otimes\left(C_{i,k}^{s_{i}^{\prime}}\right)^{\ast}\label{eq:MPDO}
\end{equation}
where each $C_{i,k}^{s_{i}}$ is a square matrix of dimension $\chi=\sqrt{\eta}$, which is often known as the \emph{bond dimension} of an MPS. The upper limit of the summation $\kappa$ is known as the \emph{Kraus dimension}. Physically, we can view the state formed by $\{C_{i,k}^{s_{i}}\}$ as an MPS for a system where each qubit is  coupled to an auxiliary qudit of dimension $\kappa$. The MPO formed by $\{A_{i}^{s_{i}s_{i}^{\prime}}\}$ is then a reduced state of this MPS for the subsystem of $N$ qubits and is therefore always physical. An MPO satisfying Eq.\,\eqref{eq:MPDO} is also called a matrix product density operator (MPDO) \cite{verstraete_matrix_2004} or a 1D locally purified tensor network (LPTN) \cite{werner_positive_2016}.

When the target state is a pure state $|\psi\rangle=\sum_{\bm{s}}\psi_{\bm{s}}|\bm{s}\rangle$, the MPDO given by Eq.\,\eqref{eq:MPDO} reduces to an MPS formed by $\{C_{i}^{s_{i}}\}$ with the Kraus dimension being 1 and
\begin{equation}
\psi_{\bm{s}}=\text{Tr}(C_{1}^{s_{1}}C_{2}^{s_{2}}\cdots C_{N}^{s_{N}})\label{eq:MPS}
\end{equation}
For a pure state, the MPS representation is preferred since the matrices in the MPO have a large dimension ($\eta=\chi^{2}$).

The MPS/MPO has a nice structure that any local unitary transformation on the state only changes the corresponding matrices locally. Since the local SIC-POVM measurement setup we introduced earlier only involves local unitaries, we can represent the probability $P(\bm{m})$ of getting a specific measurement outcome of the ququarts $\bm{m}=(m_{1},m_{2},\cdots,m_{N})$ with $m_{i}=0,1,2,3$ also in an MPO format:
\begin{align}
P(\bm{m}) & =\mathrm{Tr}\left(B_{1}^{m_{1}}...B_{N}^{m_{N}}\right)\label{eq:Pm}\\
B_{i}^{0} & =\frac{A_{i}^{00}}{2}\nonumber \\
B_{i}^{1} & =\frac{1}{6}\left(A_{i}^{00}+2A_{i}^{11}+\sqrt{2}A_{i}^{01}+\sqrt{2}A_{i}^{10}\right)\nonumber \\
B_{i}^{2} & =\frac{1}{6}\left(A_{i}^{00}+2A_{i}^{11}+\sqrt{2}e^{i\frac{2\pi}{3}}A_{i}^{01}+\sqrt{2}e^{-i\frac{2\pi}{3}}A_{i}^{10}\right)\nonumber \\
B_{i}^{3} & =\frac{1}{6}\left(A_{i}^{00}+2A_{i}^{11}+\sqrt{2}e^{-i\frac{2\pi}{3}}A_{i}^{01}+\sqrt{2}e^{i\frac{2\pi}{3}}A_{i}^{10}\right)\nonumber 
\end{align}

Performing the local SIC-POVM on all qubits in the target state provides unbiased sampling of the above probability distribution $P(\bm{m})$. To achieve QST, we want to estimate the matrices $\{A_{i}^{s_{i}s_{i}^{\prime}}\}$ that determine the distribution $P(\bm{m})$. This is however not an easy task because in practice we can only obtain a small number of measurement samples (at most polynomial in $N$) compared to the total number of $\bm{m}$ values (which is $4^{N}$). Therefore, the estimation is in general not perfect and it is not obvious how we can perform the most accurate estimation based on a given number of samples. Directly solving the matrices $\{A_{i}^{s_{i}s_{i}^{\prime}}\}$ from the above nonlinear equations is generally not possible for a large $N$, and an efficient classical algorithm is needed. We address these challenges in the next two sections.

\section{Cramer-Rao bounds}\label{sec:Cramer-Rao-bounds}

QST can be generally treated as a parameter estimation problem. Our goal is to estimate the density matrix (the collection of its matrix elements) for a target state $\rho_{0}$ using some model (also known as ansatz in quantum physics). Denote the estimation of $\rho_{0}$ by $\rho$, we can quantify the QST error using
\begin{equation}
\mathcal{R}=||\rho-\rho_{0}||_{2}^{2}\equiv\text{Tr}[(\rho-\rho_{0})^{2}]\label{eq:R}
\end{equation}
If we know the target state is a pure state $|\psi_{0}\rangle$, we can use a pure state $|\psi\rangle$ to approximate $|\psi_{0}\rangle$ and the above quantity $\mathcal{R}$ reduces to
\begin{equation}
\mathcal{R}=2(1-F)\label{eq:R-F}
\end{equation}
where $F\equiv|\langle\psi|\psi_{0}\rangle|^{2}$ is the fidelity between $|\psi\rangle$ and $|\psi_{0}\rangle$. If both $\rho$ and $\rho_{0}$ are mixed, $\mathcal{R}$ is usually normalized by the purity of the target state $\text{Tr}(\rho_{0}^{2})$ to avoid $\mathcal{R}$ being artificially small due to a low purity of the target state \cite{Baumgratz2013,BaumgratzMLE2013}.

We assume that the model state $\rho$ used to estimate the target state is a function of a list of independent real parameters $\bm{\theta}\equiv(\theta_{1},\cdots,\theta_{\mathcal{N}})$. In the current case of MPS/MPDO based QST, $\bm{\theta}$ is the list of all matrix elements (including both their real and imaginary parts) of $\{C_{i,k}^{s_{i}}\}$ in the MPDO representation of the state [see Eqs.\,\eqref{eq:MPO}-\eqref{eq:MPDO}]. Note that there is no constraint on the MPS/MPDO parameters since $\rho$ is guaranteed Hermitian and semi-positive definite. The normalization condition $\text{Tr}(\rho)=1$ is not necessary since one can divide the above distance metric $\mathcal{R}$ by $\text{Tr}(\rho)^2$ instead, and the quantity $\text{Tr}(\rho)$ can be computed efficiently due to the MPS/MPDO representation of $\rho$ \cite{perez-garcia_matrix_2007,verstraete_matrix_2004}. The target state $\rho_{0}$ is obtained by setting $\bm{\theta}=\bm{\theta}_{0}$. Here we assume the model state's MPDO has the same bond dimension as the target state's for simplicity. In practice, the target state is usually unknown and one need to make a guess of the model state's bond dimension. This guess can be refined by increasing or decreasing the bond dimension until a convergence of the MLE cost function is reached.

Define $\rho_{k}$ to be the $k^{\text{th}}$ element of the matrix $\rho$ in the computational basis where $k=1,2,\cdots4^{N}$ is a linearized index of $(\bm{s},\bm{s}^{\prime})$, and $\tilde{\rho}\equiv\rho-\rho_{0}$. We now invoke the Cramer-Rao bound to lower bound the covariance between $\tilde{\rho}_{k}$ and $\tilde{\rho}_{l}$ when $M$ observations of the target state is made \cite{Pintelon1996,Wang2020}:
\begin{equation}
\text{Cov}(\tilde{\rho}_{k},\tilde{\rho}_{l})\ge\frac{1}{M}\sum_{\alpha,\beta}\left.\frac{\partial\rho_{k}}{\partial\theta_{\alpha}}(I^{-1})_{\alpha\beta}\frac{\partial\rho_{l}^{\ast}}{\partial\theta_{\beta}}\right|_{\bm{\theta}=\bm{\theta}_{0}}\label{eq:CRbound}
\end{equation}
where $\alpha=1,2,\cdots,\mathcal{N}$ is an index of the parameter set $\bm{\theta}$. $\text{Cov}(\tilde{\rho}_{k},\tilde{\rho}_{l})$ denotes the statistical average of $\tilde{\rho}_{k}\tilde{\rho}_{l}$ where the statistical fluctuations are due to a finite number (denoted by $M$) of random samples used in estimating $\rho_{0}$. We note that the above Cramer-Rao bound requires $\rho$ to be an unbiased estimator of $\rho_{0}$, meaning that expectation value of $\tilde{\rho}$ should be a zero matrix, which may not be true. However, recent work \cite{Usami2003,GS2000} shows that the Cramer-Rao bound in Eq.\,\eqref{eq:CRbound} is asymptotically correct (for a large sample size $M$) as long as $\rho$ is a well-behaved estimator of $\rho_{0}$, which is indeed the case here since each element of $\rho$ is an analytic function of the parameter set $\bm{\theta}$ and that $\rho=\rho_0$ upon setting $\bm{\theta}=\bm{\theta}_{0}$.
 
In the above Eq.\,\eqref{eq:CRbound}, $I$ is the Fisher information
matrix whose elements are defined as 
\begin{equation}
I_{\alpha\beta}=\left.\sum_{\bm{m}}P(\bm{m})\frac{\partial\log P(\bm{m})}{\partial\theta_{\alpha}}\frac{\partial\log P(\bm{m})}{\partial\theta_{\beta}}\right|_{\bm{\theta}=\bm{\theta}_{0}},\label{eq:Iab}
\end{equation}
which depends on the probability $P(\bm{m})$ we are sampling from experimentally. Since $P(\bm{m})$ is a real-valued function of a set of real parameters $\bm{\theta}$, it's not hard to see that $I$ is real, symmetric, and semi-positive definite.

Noting that $\mathbb{E}(\mathcal{R})=\sum_{k}\text{Cov}(\tilde{\rho}_{k},\tilde{\rho}_{k})$
where $\mathbb{E}$ denotes the statistical average (over many different sets of measurement samples), the Cramer-Rao bound in Eq.\,\eqref{eq:CRbound} reduces to \cite{Wang2020}
\begin{align}
\mathbb{E}(\mathcal{R}) & \ge \frac{1}{M}\text{Tr}(KI^{-1})\label{eq:CRB}\\
K_{\alpha\beta} & \equiv\sum_{k}\left.\frac{\partial\rho_{k}}{\partial\theta_{\alpha}}\frac{\partial\rho_{k}^{\ast}}{\partial\theta_{\beta}}\right|_{\bm{\theta}=\bm{\theta}_{0}}\label{eq:Kab}
\end{align}
where $K$ is a square matrix of dimension $\mathcal{N}$ that is always Hermitian and semi-positive definite. Here $I^{-1}$ denotes the pseudo-inverse of $I$. A matrix inverse of $I$ usually does not exist since $P(\bm{m})$ can be independent of certain linear combinations of the parameters $(\theta_{1},\cdots,\theta_{\mathcal{N}})$, which leads to a null space of $I$. This happens when there are gauge degrees of freedom of the MPS/MPDO. For example, adding a phase on each element of a $C_{i,k}^{s_{i}}$ matrix for both $s_{i}=0$ and $s_{i}=1$ will leave the MPDO in Eq.\,\eqref{eq:MPDO} invariant, and thus $P(\bm{m})$ is independent of such phase. The pseudo-inverse on $I$ avoids this problem and allows for a proper calculation of the Cramer-Rao bound, because the $K$ matrix will share the same null space with $I$. A more physical way to see this is to note that $\text{Tr}(KI^{-1})$ bounds the distance metric $\mathcal{R}$ that has no gauge degree of freedom since $P(\bm{m})$ cannot be identical for two states $\rho$ and $\rho_{0}$ that are not identical due to information completeness.

A key contribution made in this paper is to show that the $K$ matrix can be evaluated efficiently while the $I$ matrix can be estimated efficiently when the target state $\rho_0$ takes the MPDO form in Eq.\,\eqref{eq:MPDO}. Let us first try to evaluate the matrix element $K_{\alpha\beta}$ using the structure of the MPO. For convenience, we index each real parameter $\theta_{\alpha}$ in the matrix $C_{i,k}^{s}$ by $\alpha=(i,j,k,s,\xi)$ where $j=1,2,\cdots,\chi^{2}$ denotes the linear index of the corresponding matrix element. $\xi=0$ ($1$) if $\theta_{\alpha}$ is the real (imaginary) part of the matrix element. One can then calculate $\frac{\partial\rho_{k}}{\partial\theta_{\alpha}}$ analytically as
\begin{align}
\frac{\partial\rho_{\bm{s},\bm{s}^{\prime}}}{\partial\theta_{\alpha}} & =\mathrm{Tr}\left(A_{1}^{s_{1}s_{1}^{\prime}}\cdots D_{i,\alpha}^{s_{i}s_{i}^{\prime}}\cdots A_{N}^{s_{N}s_{N}^{\prime}}\right)\label{eq:Drho}\\
D_{i,\alpha}^{s_{i}s_{i}^{\prime}} & = i^{\xi}\delta_{s_{i},s}E_{j}\otimes\left(C_{i,k}^{s_{i}^{\prime}}\right)^{\ast}+i^{-\xi}\delta_{s_{i}^{\prime},s}C_{i,k}^{s_{i}}\otimes E_{j}\nonumber 
\end{align}
where $E_{j}$ denotes a $\chi\times\chi$ matrix of all zeros except that its $j^{\text{th}}$ element is $1$.

Eq.\,\eqref{eq:Drho} is in an MPO format and we can also treat it as an MPS for $N$ ququarts. Therefore, we can evaluate $K_{\alpha\beta}=\sum_{k}\left.\frac{\partial\rho_{k}}{\partial\theta_{\alpha}}\frac{\partial\rho_{k}^{\ast}}{\partial\theta_{\beta}}\right|_{\bm{\theta}=\bm{\theta}_{0}}$ by contracting two such MPS, which can be performed efficiently on a classical computer \cite{perez-garcia_matrix_2007}.

For calculating the Fisher information matrix $I$, we first calculate the derivative $\frac{\partial P(\bm{m})}{\partial\theta_{\alpha}}$
used in $I_{\alpha\beta}$ analytically as
\begin{align}
\frac{\partial P(\bm{m})}{\partial\theta_{\alpha}} & =\mathrm{Tr}\left(B_{1}^{m_{1}}...G_{i,\alpha}^{m_{i}}\cdots B_{N}^{m_{N}}\right)\label{eq:DLP}\\
G_{i,\alpha}^{0} & =\frac{1}{2}D_{i,\alpha}^{00}\nonumber \\
G_{i,\alpha}^{1} & =\frac{1}{6}\left(D_{i,\alpha}^{00}+2D_{i,\alpha}^{01}+\sqrt{2}D_{i,\alpha}^{01}+\sqrt{2}D_{i,\alpha}^{10}\right)\nonumber \\
G_{i,\alpha}^{2} & =\frac{1}{6}\left(D_{i,\alpha}^{00}+2D_{i,\alpha}^{01}+\sqrt{2}e^{i\frac{2\pi}{3}}D_{i,\alpha}^{01}+\sqrt{2}e^{-i\frac{2\pi}{3}}D_{i,\alpha}^{10}\right)\nonumber \\
G_{i,\alpha}^{3} & =\frac{1}{6}\left(D_{i,\alpha}^{00}+2D_{i,\alpha}^{01}+\sqrt{2}e^{-i\frac{2\pi}{3}}D_{i,\alpha}^{01}+\sqrt{2}e^{i\frac{2\pi}{3}}D_{i,\alpha}^{10}\right)\nonumber 
\end{align}
which are identical to Eq.\,\eqref{eq:Pm} upon replacing $A_{i}^{s_{i}s_{i}^{\prime}}$ by $D_{i,\alpha}^{s_{i}s_{i}^{\prime}}$ and $B_{i}^{m_{i}}$ by $G_{i,\alpha}^{m_{i}}$.

While $\frac{\partial P(\bm{m})}{\partial\theta_{\alpha}}$ is also in an MPO format, we cannot evaluate $I_{\alpha\beta}$ efficiently because $I_{\alpha\beta}=\sum_{\bm{m}}\frac{1}{P(\bm{m})}\frac{\partial P(\bm{m})}{\partial\theta_{\alpha}}\frac{\partial P(\bm{m})}{\partial\theta_{\beta}}$ [see Eq.\,\eqref{eq:Iab}] contains an extra weight $\frac{1}{P(\bm{m})}$. However, we can estimate $I_{\alpha\beta}$ using Monte Carlo sampling by using the fact that both $P(\bm{m})$ and $\frac{\partial\log P(\bm{m})}{\partial\theta_{\alpha}}=\frac{1}{P(\bm{m})}\frac{\partial P(\bm{m})}{\partial\theta_{\alpha}}$ can be computed efficiently for a given $\bm{m}$. More interestingly, we can perform direct sampling over the probability distribution $P(\bm{m})$ instead of the commonly used Metropolis-Hastings algorithm in variational Monte Carlo method that may suffer from poor sampling quality. This direct sampling is achieved by utilizing a special nature of the MPO representation of $P(\bm{m})$ similar to autoregressive artificial neural networks \cite{sharir_deep_2020}. In particular, we can calculate the following series of conditional probabilities efficiently:
\begin{align}
P(m_{1}) & =\mathrm{Tr}\left(B_{1}^{m_{1}}\mathcal{B}_{2}...\mathcal{B}_{N}\right)\label{eq:Pcond}\\
P(m_{2}|m_{1}) & =\mathrm{Tr}\left(B_{1}^{m_{1}}B_{2}^{m_{2}}...\mathcal{B}_{N}\right)\nonumber \\
\vdots & \vdots\nonumber \\
P(m_{N}|m_{N-1}\cdots m_{1}) & =\mathrm{Tr}\left(B_{1}^{m_{1}}...B_{N}^{m_{N}}\right)\nonumber 
\end{align}
where we defined $\mathcal{B}_{i}=\sum_{m_{i}}B_{i}^{m_{i}}$. We can then perform the sampling of $P(\bm{m})$ efficiently (which we call autoregressive sampling) as follows:
\begin{enumerate}

\item Calculate $P(m_{1})$ for $m_{1}=0,1,2,3$, then sample from this distribution [normalized by $\sum_{m_{1}}P(m_{1})$ if $P(\bm{m})$ is not normalized] to get a random value of $m_{1}\in\{0,1,2,3\}$.

\item Depending on the value of $m_{1}$ sampled in the previous step, calculate $P(m_{2}|m_{1})$ for $m_{2}=0,1,2,3$ and sample from this distribution (again normalized) to get a random value of $m_{2}$.

\item Iterate this process to determine $m_{3},\cdots,m_{N}$.

\end{enumerate}
Note that the above process can be done in parallel to create many samples of $\bm{m}=(m_{1},\cdots,m_{N})$ efficiently on a classical computer. We then compute $\frac{\partial\log P(\bm{m})}{\partial\theta_{\alpha}}\frac{\partial\log P(\bm{m})}{\partial\theta_{\beta}}$ for all $\alpha,\beta$ indices with each sampled $\bm{m}$ and average the obtained Fisher information matrix $I$ over all samples to get an estimation of $I$. We increase the number of samples $\mathcal{M}$ until $I$ converges. For most physical target states we have studied, we find $\mathcal{M}=10^4-10^5$ sufficient to ensure accurate estimation of $I$ for $N\le40$ qubits, and the calculation of the entire $I$ matrix takes only a few minutes on a modern PC for a small bond dimension $\chi$.

With both $K$ and $I$ efficiently computable, we can now compute $\text{Tr}(KI^{-1})$ in the Cramer-Rao bound [Eq.\,\eqref{eq:CRB}] for a variety of different target states. For simplicity, we focus on pure target states as exact MPS representations exist for a variety of many-body entangled states. This also allows us to better interpret the Cramer-Rao bound using the structure of entanglement, since characterizing entanglement is much harder for a mixed state. Moreover, for a pure state we can directly lower bound the infidelity ($1-F$) of the QST using $\text{Tr}(KI^{-1})$ and the sample size $M$ using Eq.\,\eqref{eq:R-F}:
\begin{equation}
\mathbb{E}(1-F)\ge\text{\ensuremath{\frac{1}{2M}}Tr}(KI^{-1}) \label{eq:infid}
\end{equation}

\begin{figure*}
\includegraphics[width=0.33\textwidth]{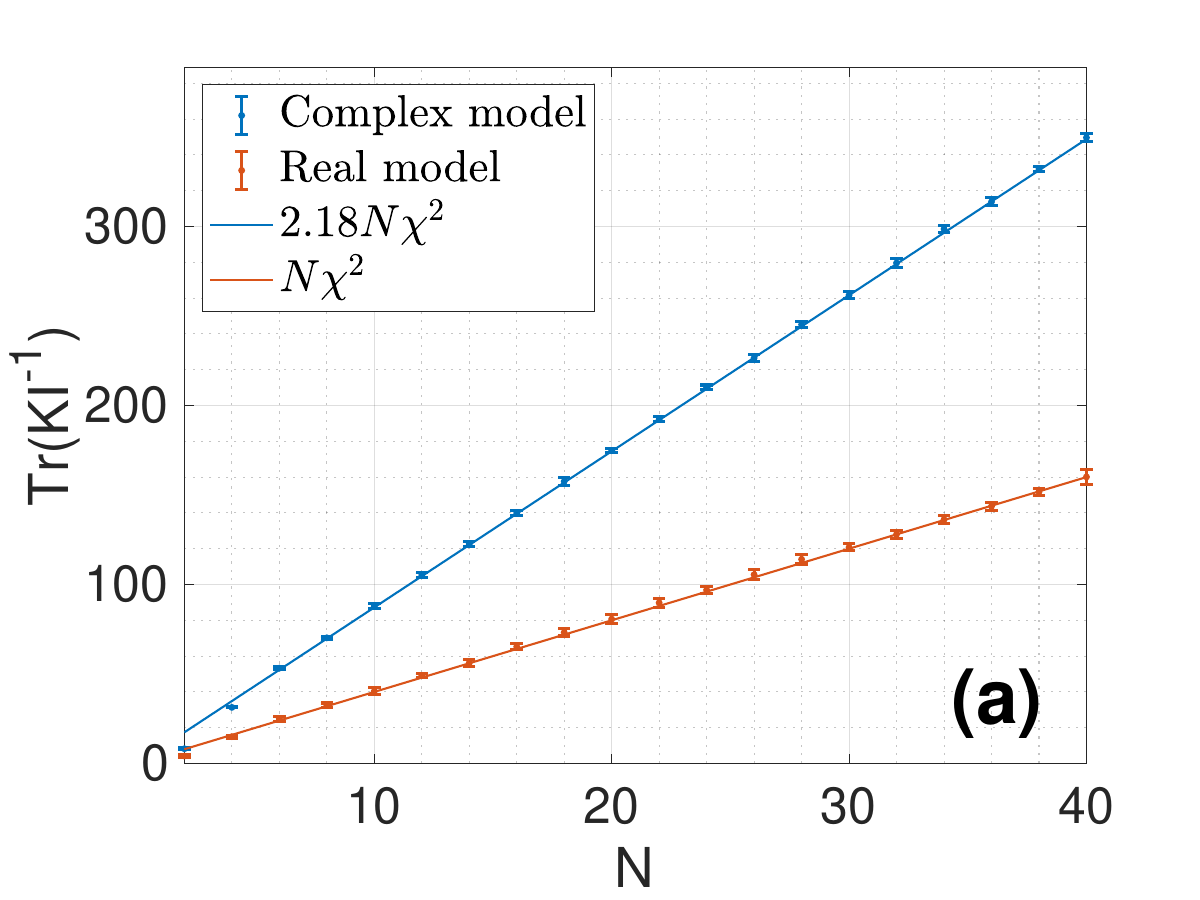}\hfill{} \includegraphics[width=0.33\textwidth]{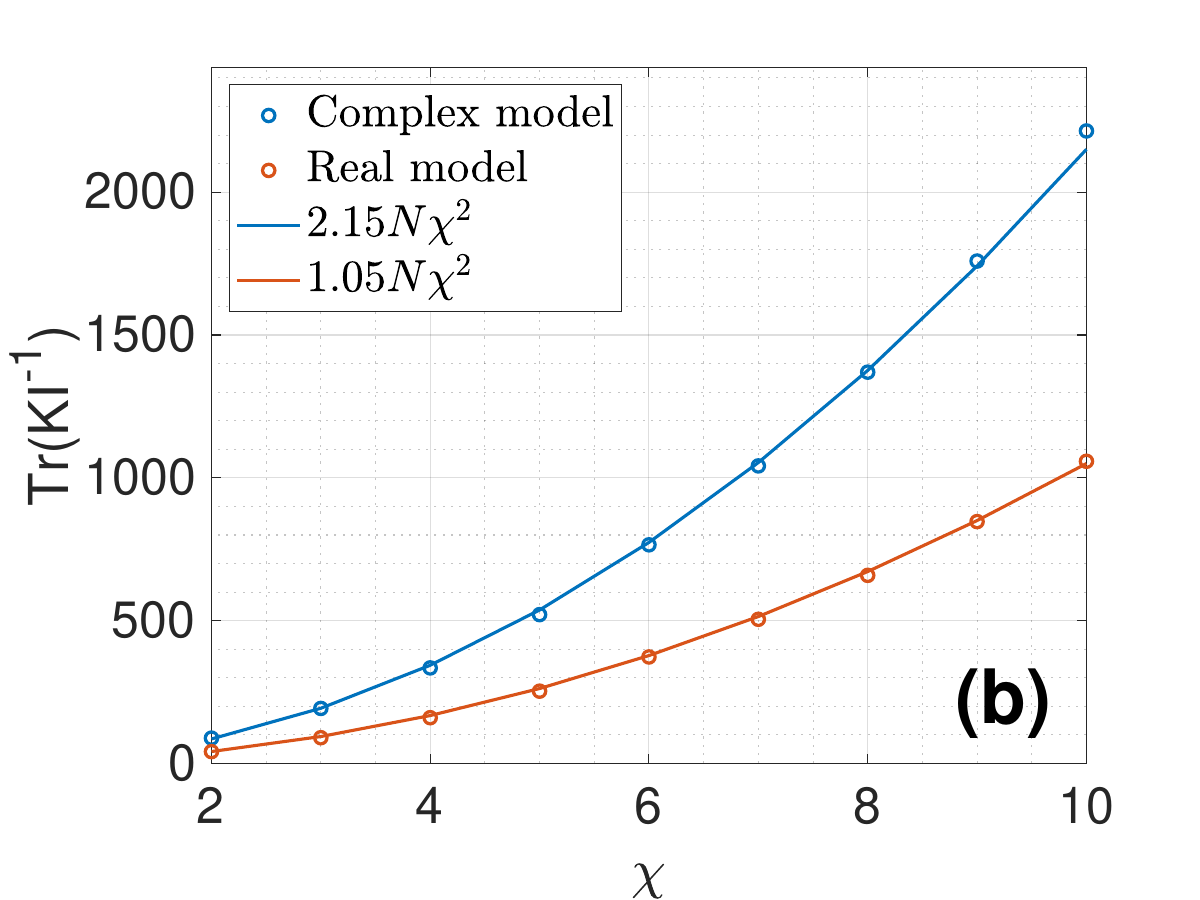}\hfill{} \includegraphics[width=0.33\textwidth]{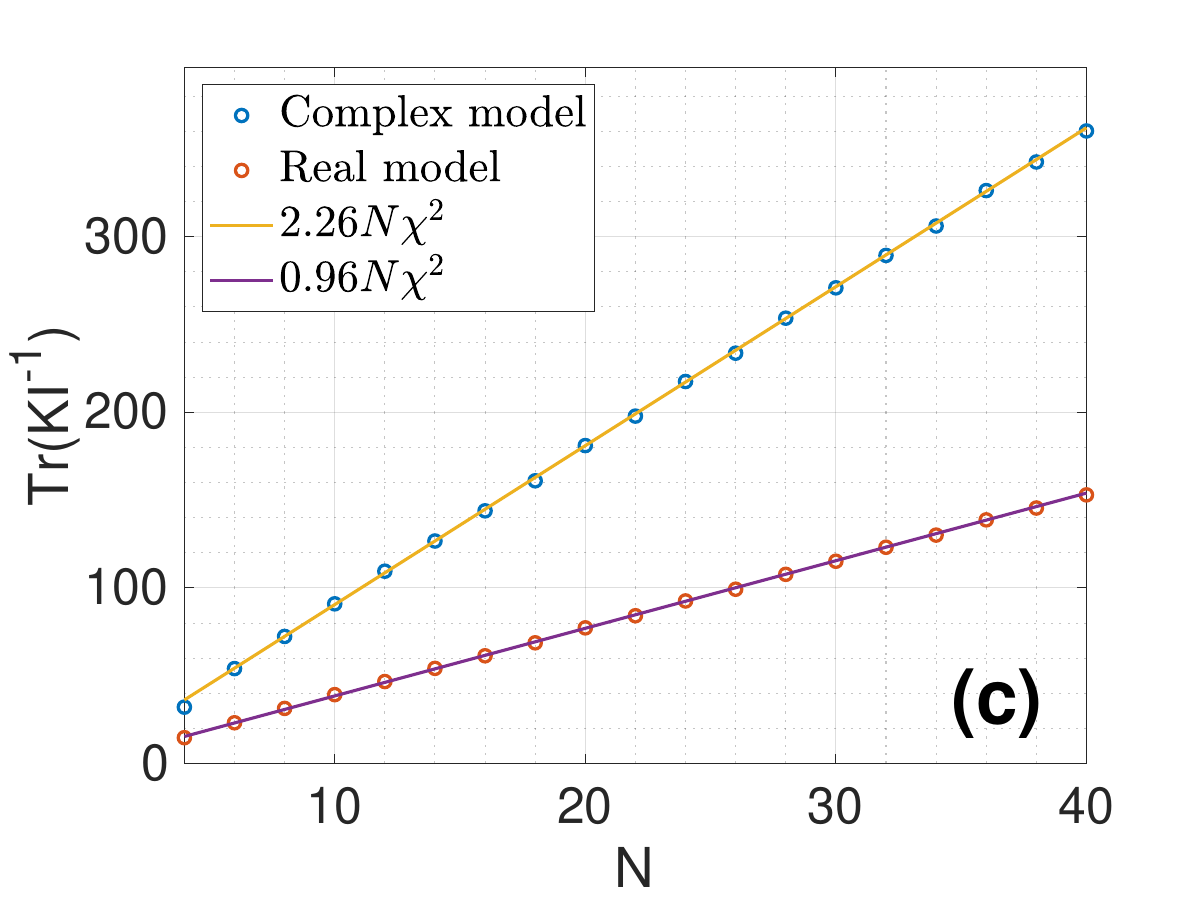}
\caption{$\text{Tr}(KI^{-1})$ calculated for the random MPS target states as a function of $N$ for $\chi=2$ (a), as a function of $\chi$ for $N=10$ (b), and for the 1D cluster state (c). We calculate $\text{Tr}(KI^{-1})$ for $10$ different random MPS target states and show their mean and standard deviation (as error bars) in (a). For each target state, we use a model MPS state with the same bond dimension and assume the matrices in the model state to be either real (blue) or complex (red). The solid lines in each plot are linear fits to the calculated $\text{Tr}(KI^{-1})$ values. $\mathcal{M}=10^{5}$ samples is used for the Monte Carlo sampling calculation of the Fisher information matrix. Further increasing $\mathcal{M}$ does not lead to visible differences for all the plots. \label{fig1}}
\end{figure*}

We start by calculating $\text{Tr}(KI^{-1})$ for a set of random MPS states where the target state is expressed by Eq.\,\eqref{eq:MPS} with each $C_{i}^{s_{i}}$ being a random $\chi\times\chi$ real matrix. The elements of each random matrix is drawn uniformly between $-1$ and $1$. The model state is also an MPS with bond dimension $\chi$, and we allow each matrix element in the MPS to be an independent parameter that is either real or complex, such that there are in total $2N\chi^{2}$ or $4N\chi^{2}$ real parameters. We compute $\text{Tr}(KI^{-1})$ for $N$ up to $40$ and $\chi$ up to $10$. The results shown in Fig.\,\ref{fig1}(a) indicate that $\text{Tr}(KI^{-1})\approx N\chi^{2}$ for a real MPS model and $\text{Tr}(KI^{-1})\approx2.2N\chi^{2}$ for a complex MPS model. 

It may be counter-intuitive that the complex MPS model leads to a larger Cramer-Rao bound than the real MPS model for the same target state, as it appears that the complex model is just an over-parameterization. However, we note that the extra parameters in the complex model are actually fully identifiable by our informationally-complete measurements, in contrast to the gauge degrees of freedom which are not identifiable. Therefore, these extra parameters will contribute to the Cramer-Rao bound and actually increase the statistical fluctuations for the estimation of the target state with a finite number of measurement samples. In both the real and the complex cases, we find that $\text{Tr}(KI^{-1})$ is proportional to the number of independent parameters describing the target state, consistent with a rough estimate of $\text{Tr}(KI^{-1})$ obtained in Ref.\,\cite{Wang2020} for random local basis measurement (which should be similar to the local SIC-POVM when the number of random bases is large). Therefore, for random MPS states, we expect efficient QST using the local SIC-POVM.

Next, we change our target state to a 1D cluster state, which is a typical short-range entangled state \cite{chen_local_2010}. This state is translationally invariant (TI) and described by the following MPS of bond dimension $\chi=2$: \cite{Perez2007}
\begin{equation}
C_{\text{cluster}}^{0}=\begin{pmatrix}0 & 0\\
1 & 1
\end{pmatrix}\quad C_{\text{cluster}}^{1}=\begin{pmatrix}1 & -1\\
0 & 0
\end{pmatrix}\label{eq:cluster}
\end{equation}
Physically, this state can be generated by applying controlled-Z gates on all neighboring pairs of qubits on a 1D ring starting from an initial product state of all qubits in the state $\frac{|0\rangle+|1\rangle}{\sqrt{2}}$. It is short-range entangled as it can be generated by a single-layer local quantum circuit. 

As shown by Fig.\,\ref{fig1}(c), we find that for the cluster state, $\text{Tr}(KI^{-1})$ scales linearly in $N$ with the complex model again having its value roughly twice of the real model, similar to the results for the random MPS state [see Fig.\,\ref{fig1}(a-b)]. Note that the linear in $N$ scaling here is due to the fact that the model state is not TI and has $\mathcal{O}(N)$ parameters that are identifiable by the measurements. A TI model would lead to $\text{Tr}(KI^{-1})$ independent of $N$. As a result, we see that the number of state copies needed to reconstruct a given target state within a given error  depends on how many parameters are used in the model. We can use less number of parameters if we have more \emph{a priori} knowledge of the target state, such as efficient MPS/MPO representation, real vs complex wavefunctions, TI vs non-TI, etc. In other words, the knowledge we have for the target state leads to a constraint on the estimation problem, which can significantly lower the Cramer-Rao bound \cite{Gorman1990}.

We now consider a paradigmatic example of a long-range entangled state, a GHZ state $|\psi_{\text{GHZ}}\rangle=|00\cdots0\rangle+|1\cdots1\rangle$ (not normalized for convenience). This state requires a local quantum circuit of depth $\mathcal{O}(N)$ to be created from a product state. It is described by a TI MPS with
\begin{equation}
C_{\text{GHZ}}^{0}=\begin{pmatrix}1 & 0\\
0 & 0
\end{pmatrix},\quad C_{\text{GHZ}}^{1}=\begin{pmatrix}0 & 0\\
0 & 1
\end{pmatrix}\label{eq:GHZ}
\end{equation}
Since both matrices are diagonal, we can in fact calculate the $K$ and $I$ matrices semi-analytically with either a TI or non-TI MPS model. We defer the detailed analytical calculations to the Appendix A, and summarize the results below.

\begin{figure*}
\includegraphics[width=0.33\textwidth]{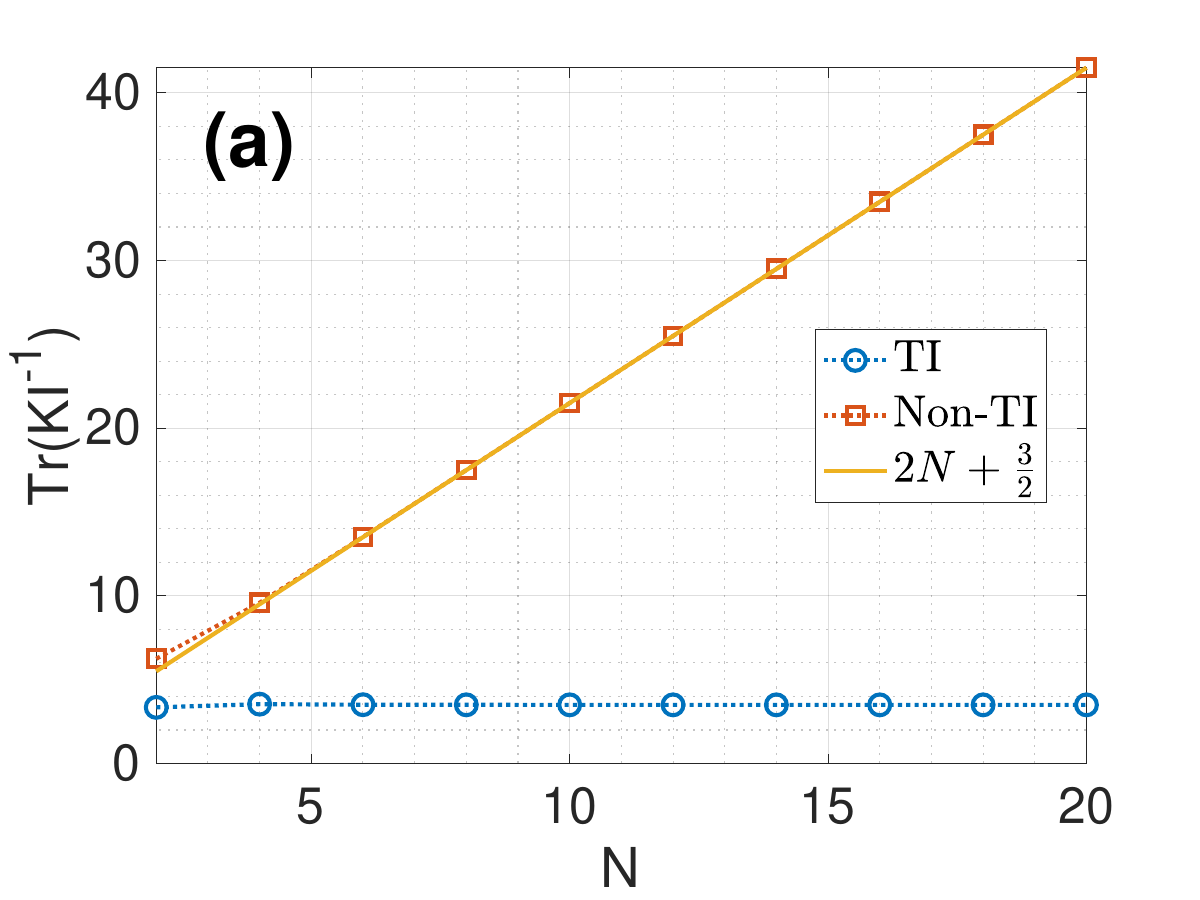}\hfill{}\includegraphics[width=0.33\textwidth]{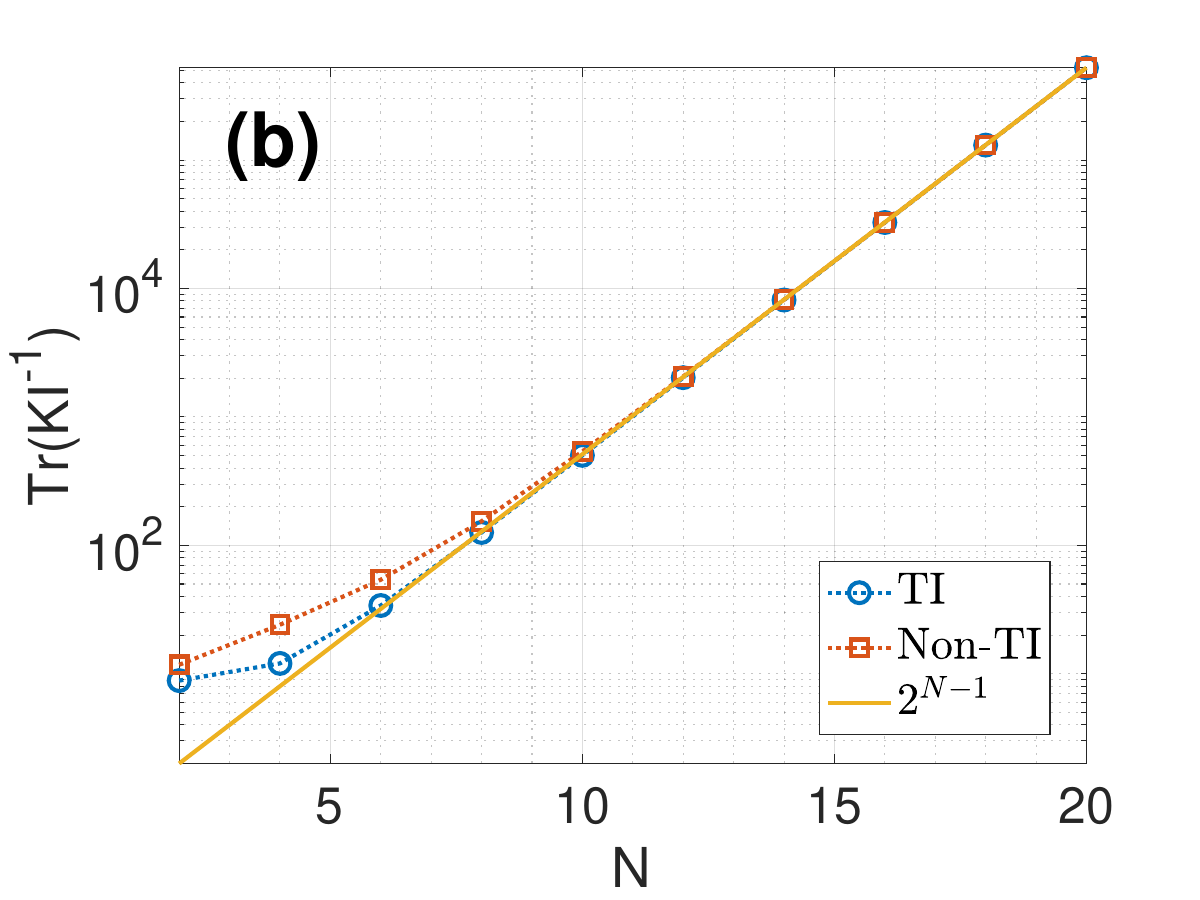}\hfill{}\includegraphics[width=0.33\textwidth]{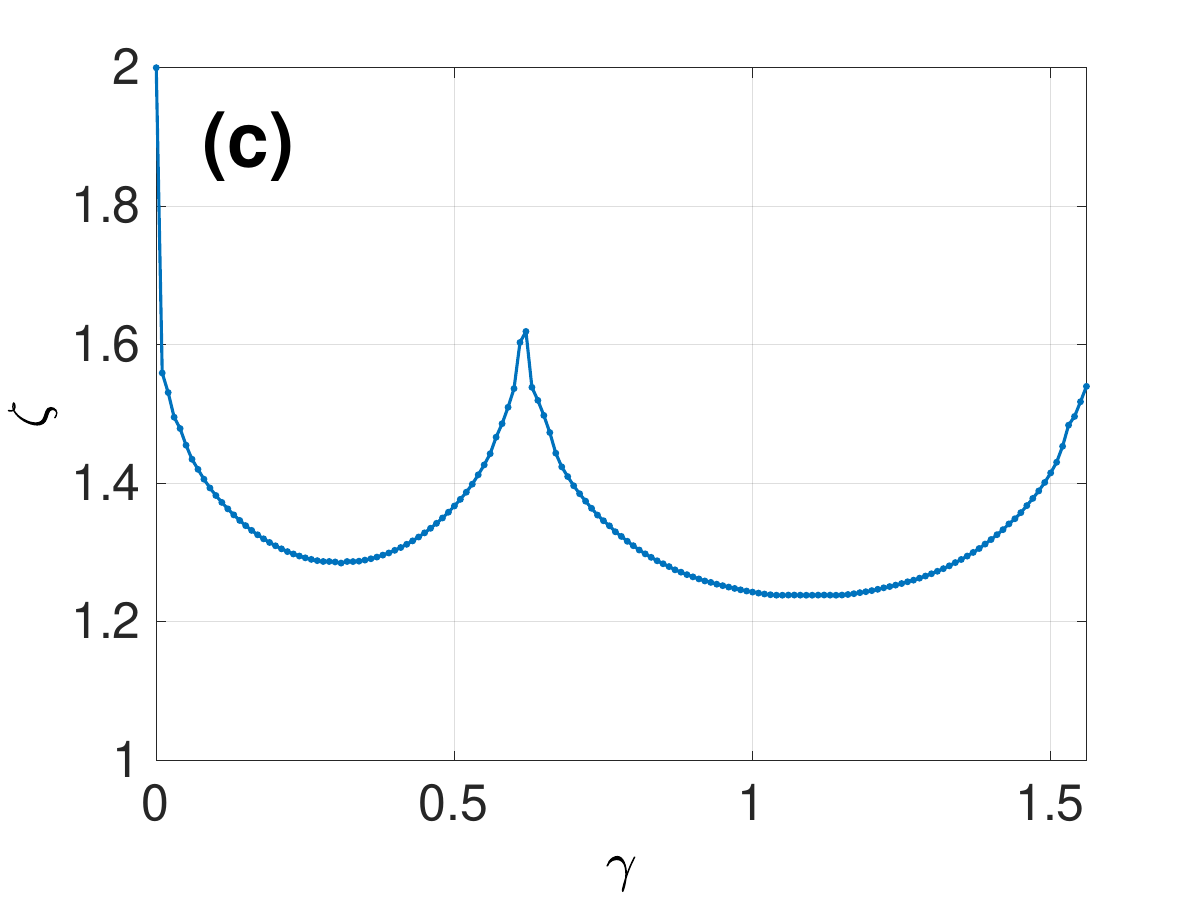} 
\caption{(a) $\text{Tr}(KI^{-1})$ calculated semi-analytically using both a translationally invariant (TI) and a non-translationally invariant (Non-TI) real MPS model for the GHZ state in Eq.\,\eqref{eq:GHZ}. (b) Same as (a) but with a complex MPS model. (c) $\text{Tr}(KI^{-1})$ for the generalized GHZ state in Eq.\,\eqref{eq:gGHZ} is found to be proportional to $\zeta^{N}$ for large $N$. The plot shows the value of $\zeta$ as a function of the rotation angle $\gamma$ obtained via a fit of the semi-analytically computed $\text{Tr}(KI^{-1})$ values for $N=20,22,\cdots,60$.
\label{fig2}}
\end{figure*}

If we assume the model MPS contains only real parameters, then $\text{Tr}(KI^{-1})$ quickly converges to $\frac{7}{2}$ for a TI model and $2N+\frac{3}{2}$ for a non-TI model [see Fig.\,\ref{fig2}(a)]. However, if we assume each matrix element can be complex, then $\text{Tr}(KI^{-1})$ quickly converges to $2^{N-1}$ for both a TI and a non-TI model [see Fig.\,\ref{fig2}(b)]. The real MPS model behaves similarly to the target states we studied previously, where $\text{Tr}(KI^{-1})$ scales with the number of parameters in the model. The complex MPS model leads to a sharply different scenario, with $\text{Tr}(KI^{-1})$ scales exponentially in $N$. 

This exponential scaling is due to the difficulty in determining the relative phase between the two component states $|0\cdots0\rangle$ and $|1\cdots1\rangle$. To illustrate this, we consider a minimal model for estimating the GHZ state with just two real parameters $\phi_{1}$ and $\phi_{2}$:
\begin{equation}
C_{\phi_{1}}^{0}=\begin{pmatrix}e^{i\phi_{1}} & 0\\
0 & 0
\end{pmatrix},\quad C_{\phi_{2}}^{1}=\begin{pmatrix}0 & 0\\
0 & e^{i\phi_{2}}
\end{pmatrix}\label{eq:mGHZ}
\end{equation}
It's easy to see that this model represents a modified GHZ state $|\phi_{\text{GHZ}}\rangle=e^{iN\phi_{1}}|0\cdots0\rangle+e^{iN\phi_{2}}|1\cdots1\rangle$. The analytical calculations described in Appendix A allow us to find the $K$ and $I$ matrices explicitly:
\begin{equation}
K=\frac{N^{2}}{2}\begin{pmatrix}1 & -1\\
-1 & 1
\end{pmatrix},\quad I=N^{2}\delta\begin{pmatrix}1 & -1\\
-1 & 1
\end{pmatrix}\label{eq:KI-GHZ}
\end{equation}
where $\delta\approx\frac{1}{2^{N}}$ and $\text{Tr}(KI^{-1})\approx2^{N-1}$ (asymptotically exact in the large $N$ limit) , same as that using a general complex MPS model with $\chi=2$ (see Fig.\,\ref{fig2}(b) and Appendix A). There are two shared eigenvectors of $K$ and $I$ in Eq.\,\eqref{eq:KI-GHZ}. One has eigenvalue $0$, corresponding to $\phi_{1}+\phi_{2}$ which only changes the global phase of the state $|\phi_{\text{GHZ}}\rangle$. The other has eigenvalue $\delta\approx2^{-N}$ for the $I$ matrix, corresponding to $\phi_{1}-\phi_{2}$, which controls the relative phase between the two component states in $|\psi_{\text{GHZ}}\rangle$. An exponentially small eigenvalue of the Fisher information matrix denotes that the measurement we performed gives little information for $\phi_{1}-\phi_{2}$, and thus requires an exponentially large sample size in $N$ to obtain a good estimate.

The real MPS model cannot estimate this relative phase and hence does not lead to an exponential scaling in the Cramer-Rao bound. The real MPS model implies a constraint on the target state, requiring it to be represented only by a real MPS. While such a constraint is expected to lower the Cramer-Rao bound, it is surprising that this constraint leads to an exponentially smaller bound. It should also be pointed out that constraining the target state to have real wavefunctions in the computational basis (up to arbitrary single-qubit rotations) does not make QST easy or trivial since there are many interesting and complicated many-body entangled states that are real. For example, the eigenstates of any real Hamiltonian are real. Therefore, the real MPS model used to achieve efficient QST for the GHZ state is also capable of reconstructing any quantum many-body state represented by a real MPS (noting that the bond dimension of the MPS model can be increased without changing the Cramer-Rao bound for the GHZ state). We thus argue that local reducibility is not necessary for efficient QST using MPS/MPO, since a real GHZ state is not locally reducible. We conjecture that efficient QST is possible for any target state with an efficient real MPS representation.

Finally, we have also considered a generalized GHZ state $|\Psi_{\text{GHZ}}\rangle=U_{\gamma}^{\otimes N}|\psi_{\text{GHZ}}\rangle$ that differs from the standard GHZ state by a global SO(2) rotation, where $U_{\gamma}=\begin{pmatrix}\cos\gamma & -\sin\gamma\\
\sin\gamma & \cos\gamma
\end{pmatrix}$. $|\Psi_{\text{GHZ}}\rangle$ is represented by the TI-MPS
\begin{equation}
C_{\gamma}^{0}=\begin{pmatrix}\cos\gamma & 0\\
0 & \sin\gamma
\end{pmatrix},\quad C_{\gamma}^{1}=\begin{pmatrix}-\sin\gamma & 0\\
0 & \cos\gamma
\end{pmatrix}\label{eq:gGHZ}
\end{equation}

Interestingly, a complex MPS model now yields $\text{Tr}(KI^{-1})\sim\zeta^{N}$ where the base of the exponential $\zeta$ depends on the rotation angle $\gamma$. As shown in Fig.\,\ref{fig2}(c), the original GHZ state (where $\gamma=0$) in fact has the largest $\zeta$ ($\zeta_{\text{max}}=2$), while $\zeta_{\text{min}}\approx1.24$ is achieved for $\gamma\approx\frac{\pi}{3}$. This implies that there also exists an optimal local SIC-POVM, one that differs from our choice [Eq.\,\eqref{eq:SIC-POVM}] by the local unitary transformation $U_{\gamma}$. While we believe that the exponential scaling of $\text{Tr}(KI^{-1})$ applies to any local SIC-POVM, an optimal choice could lead to a much smaller infidelity for QST. However, finding such optimal measurement setting would require \emph{a priori} knowledge of the target state and is not preferred for a universal QST protocol.

\section{MLE algorithm and benchmark results}\label{ExpResultsSection}

The Cramer-Rao bound studied in the previous section provides a lower bound on the expectation value of the QST infidelity (for pure states) for a given number of state copies and a specific target state. Whether this lower bound can be achieved is a separate question. It is well known that the MLE algorithm can saturate the Cramer-Rao bound \cite{Braunstein1992} theoretically in the large sample size limit. However, in practice the MLE algorithm may not be able to find the global optimum when the number of parameters is large and thus the saturation of the Cramer-Rao bound is not guaranteed.

Here we describe the MLE algorithm we used and benchmark its performance using synthetic experimental data and compare it with the Cramer-Rao bound calculated using the method developed in the previous section. Specifically, we define a cost function that is the negative log-likelihood over all $M$ samples obtained during the local SIC-POVM measurements for the model MPS/MPDO state:
\begin{equation}
    D_{\text{NLL}}  =-\frac{1}{M}\sum_{(m_{1},...,m_{N})\in\mathcal{V}}\log\left(\mathrm{Tr}[B_{1}^{m_{1}}\cdots B_{N}^{m_{N}}]\right)\label{NLL}
\end{equation}
where $\mathcal{V}$ is the set of all $M$ samples (each sample is a $N$-quart string representing a measurement outcome). Our goal is to minimize $D_{\text{NLL}}$ by optimizing the matrices $\{C_{i,k}^{s_{i}}\}$ in the model state with a guessed bond dimension and Kraus dimension. The matrices $\{B_{i}^{m_{i}}\}$ in Eq.\,\eqref{NLL} are determined from $\{C_{i,k}^{s_{i}}\}$ using Eq.\,\eqref{eq:MPDO} and Eq.\,\eqref{eq:Pm}.

The optimization of a cost function such as the $D_{\text{NLL}}$ above that depends on many parameters in a complicated, nonlinear way is a central goal in modern machine learning tasks. Here we use the well-known PyTorch library to perform such optimization with a stochastic gradient descent (SGD) optimizer or an Adam optimizer. To better achieve the global minimum of the cost function $D_{\text{NLL}}$, we have also used several different methods for scheduling the learning rate of the optimizer, including a Nesterov momentum method and a cosine annealing scheduler. In addition, each optimization is performed over multiple random initial guesses of the matrices $\{C_{i,k}^{s_{i}}\}$, and we pick the run with the lowest converged cost function $D_{\text{NLL}}$ and then calculate the distance metric $\mathcal{R}$ [see in Eq.\,\eqref{eq:R}] between the corresponding model state and the target state. We note that $\mathcal{R}$ can be computed efficiently as long as the model state and target state are both efficiently represented using MPS/MPO.

The input data to the optimizer is the sample set $\mathcal{V}$ which can be generated experimentally. Here we generate the sample set synthetically for a given MPS/MPDO target state using the autoregressive sampling method described in Section \ref{sec:Cramer-Rao-bounds}. Since the sample size $M$ is finite, we shall repeat the entire optimization for multiple sets (10 for each of the calculations below) of random samples with the same size and average over the calculated $\mathcal{R}$ to obtain $\mathbb{E}(\mathcal{R})$.

\begin{figure*}
\includegraphics[width=0.33\textwidth]{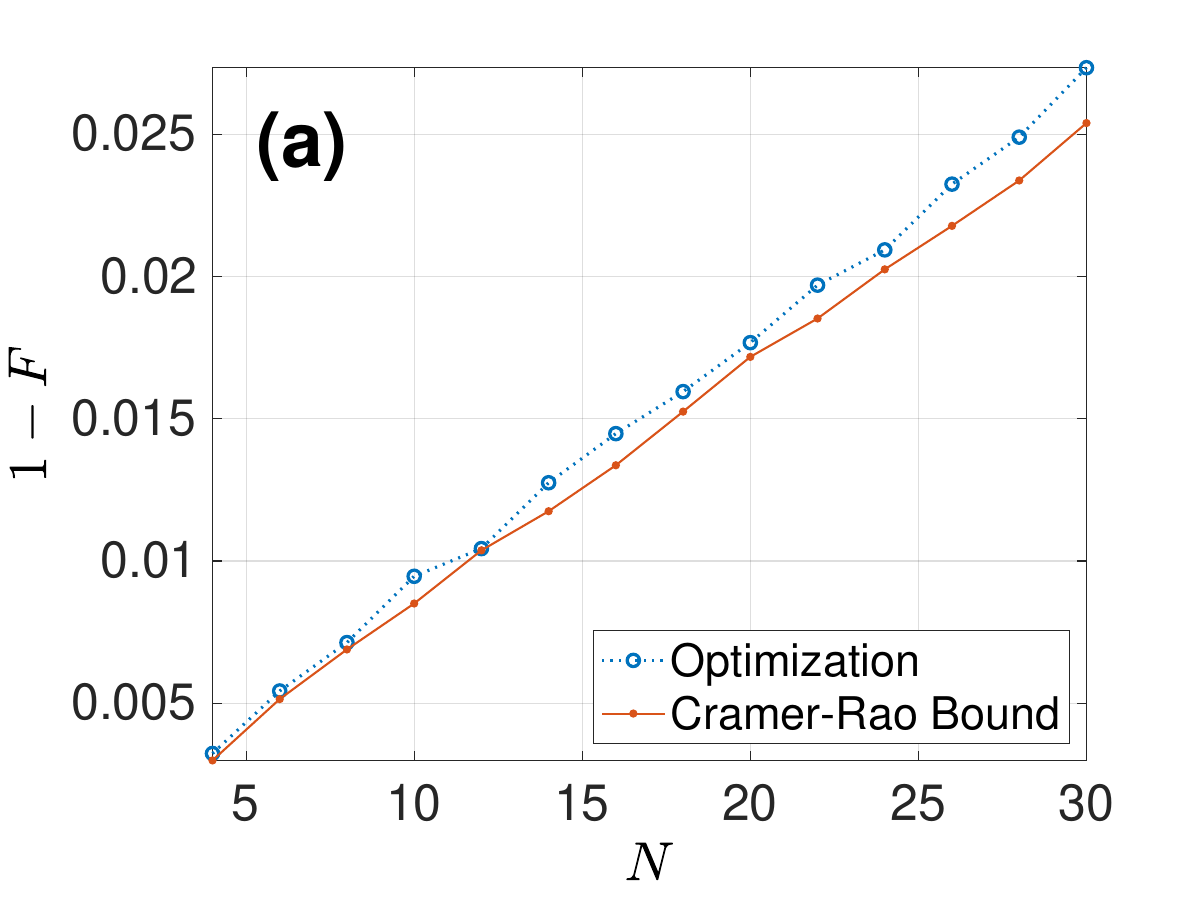} \includegraphics[width=0.33\textwidth]{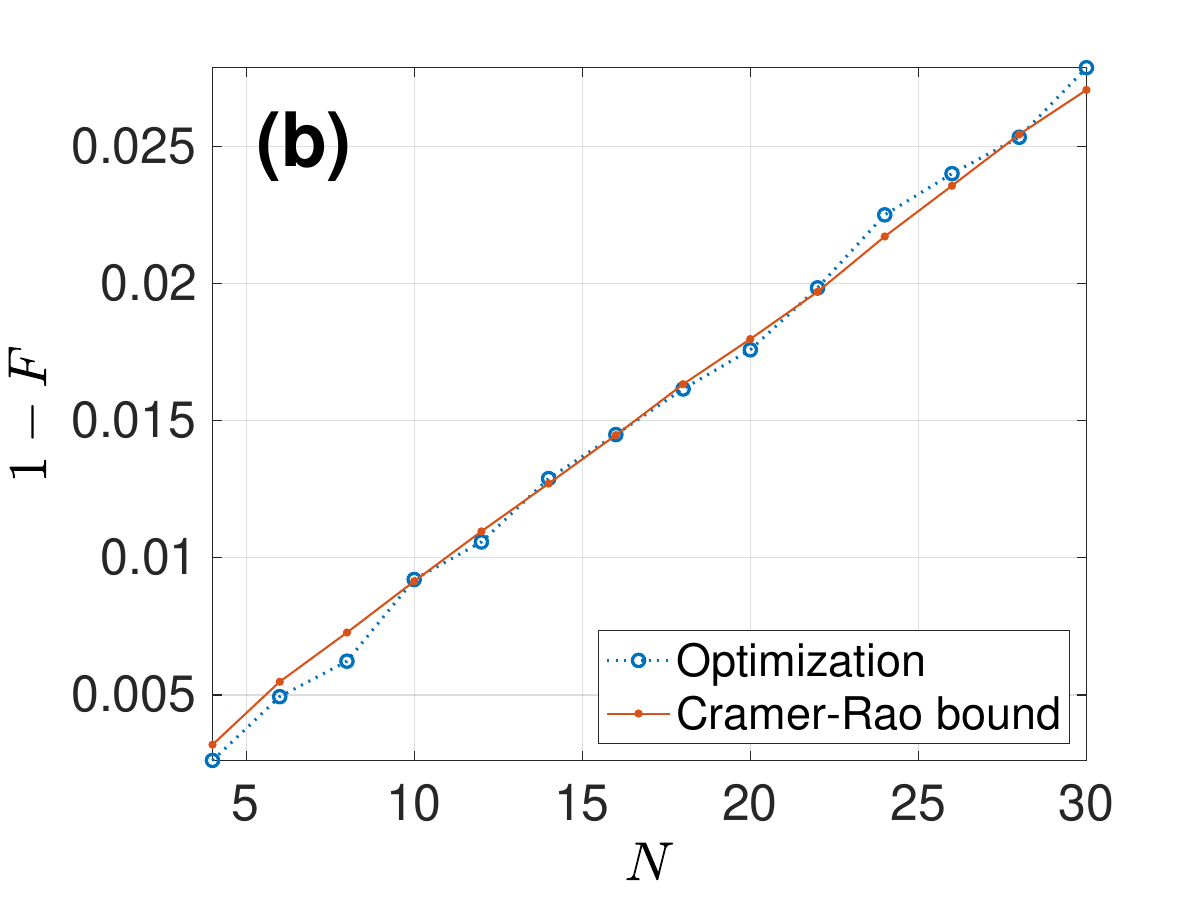} \includegraphics[width=0.33\textwidth]{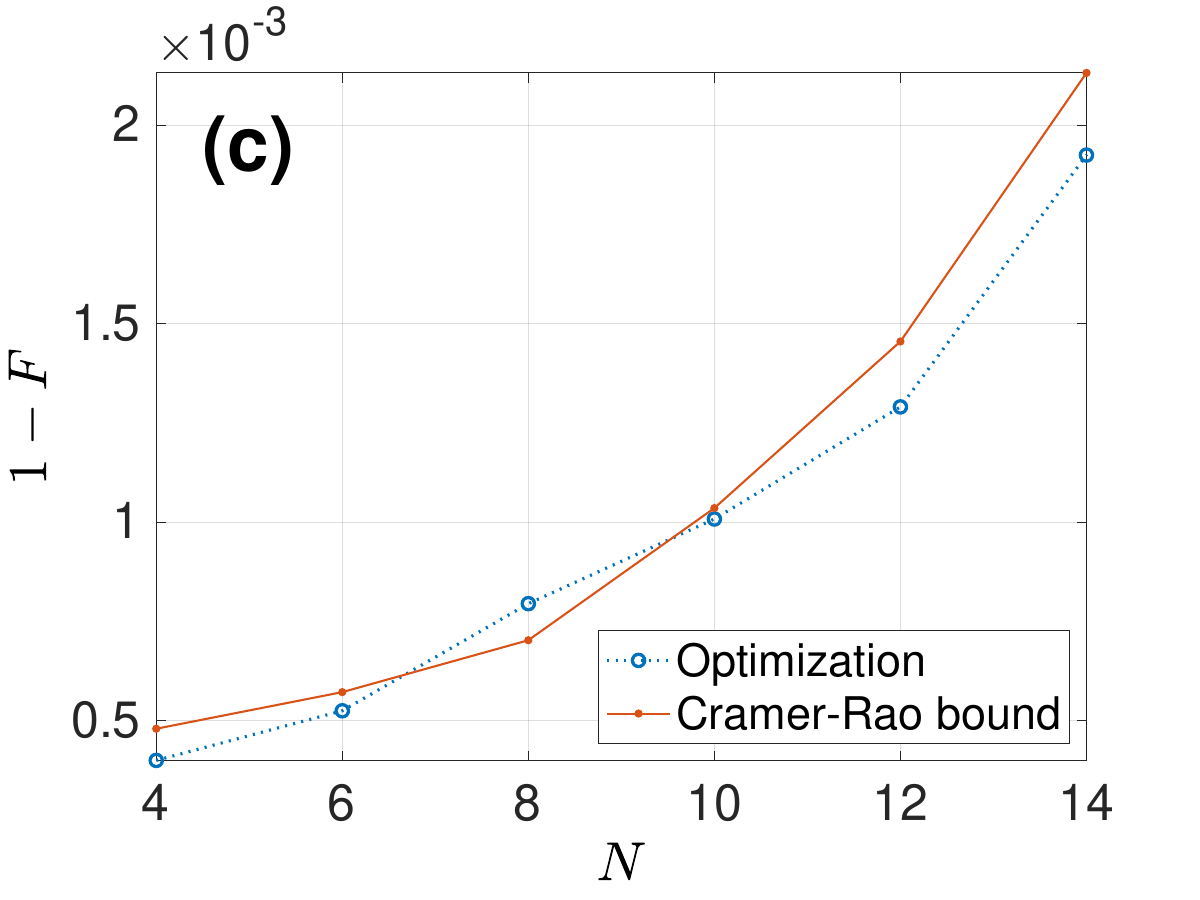}
\caption{Results of the MLE algorithm and their comparison with the Cramer-Rao bound for (a) a 1D random complex MPS state with a bond dimension $\chi=2$ and a sample size $M=5\times10^3$, (a) a 1D cluster state defined by Eq.\,\eqref{eq:cluster} with a complex MPS model and a sample size $M=5\times10^3$, and (c) a generalized GHZ state with $\gamma=\pi/3$, defined by Eq.\,\eqref{eq:gGHZ}, with a complex TI MPS model and a sample size $M=10^4$. The model MPS has a bond dimension $\chi=2$ for all target states, and the specified sample size is used for both the MLE algorithm and the prediction of QST infidelity based on the Cramer-Rao bound. \label{fig3}}
\end{figure*}

We first benchmark our MLE algorithm using the random MPS state, the 1D cluster state, and a generalized GHZ state as target states. For the random MPS state, we assume both the target state and the model state to be made of complex matrices of bond dimension $2$. For the cluster state, we use a non-TI complex MPS model of bond dimension $2$, and for the generalized GHZ state, we use a complex TI MPS model of bond dimension $2$, with the target state given by Eq.\,\eqref{eq:gGHZ} with the rotation angle $\gamma=\frac{\pi}{3}$. In all three cases, we see that the infidelity between the target state and the state obtained via the MLE algorithm closely match the minimum infidelity predicted by the Cramer-Rao bound [see Eq.\,\eqref{eq:infid}], as shown in Fig.\,\ref{fig3}. Small differences between the two are likely due to statistical fluctuations (since we have only performed the MLE algorithm for $10$ different sets of random measurement samples) and limited sample size (since the MPS/MPDO model is likely not an unbiased estimator and the Cramer-Rao bound could be inaccurate if the sample size is not large enough).

We have also benchmarked our MLE algorithm using mixed target states. The infidelity between the target state $\rho_{0}$ and the state reconstructed by the MLE algorithm (denoted by $\rho$) cannot be computed efficiently, but we can compute the following distance metric between the two states
\begin{align}
\mathcal{D}=\lVert\rho-\rho_{0}\rVert_{2}^{2}/\lVert\rho_{0}\rVert_{2}^{2} & =\mathcal{R}/\text{Tr}(\rho_{0}^{2})\label{eq:D},
\end{align}
which is the aforementioned distance metric $\mathcal{R}$ normalized by the purity of the target state that can also be computed efficiently due to the efficient MPO representations of $\rho_{0}$.

We first test our MLE algorithm for reconstructing random MPDO states where each $\{C_{i,k}^{s_{i}}\}$ matrix in Eq.\,\eqref{eq:MPDO} is a $\chi\times\chi$ random complex matrix. We choose a model MPDO state with the same bond dimension and Kraus dimension as the target state. Second, we choose target states that are thermal states of the following 1D quantum Ising model
\begin{align}
H_{\text{I}}= & \sum_{j=1}^{N-1}\sigma_{j}^{z}\sigma_{j+1}^{z}+B\sum_{j=1}^{N}\sigma_{j}^{x}\label{QuantumIsing}.
\end{align}
We set $B=1$, which makes the model at its quantum criticality in the thermodynamic limit. The thermal state is defined as $\rho_{T}=\frac{e^{-H_{\text{I}}/T}}{\mathrm{Tr}(e^{-H_{\text{I}}/T})}$. To obtain $\rho_{T}$ for a large system size, we use an imaginary time TEBD method \cite{jaschke_one-dimensional_2018} to approximate $\rho_{T}$ using an MPO. The MLE algorithm is then used to reconstruct the target MPO state as an MPDO. 

\begin{figure}
\centering \includegraphics[width=0.4\textwidth]{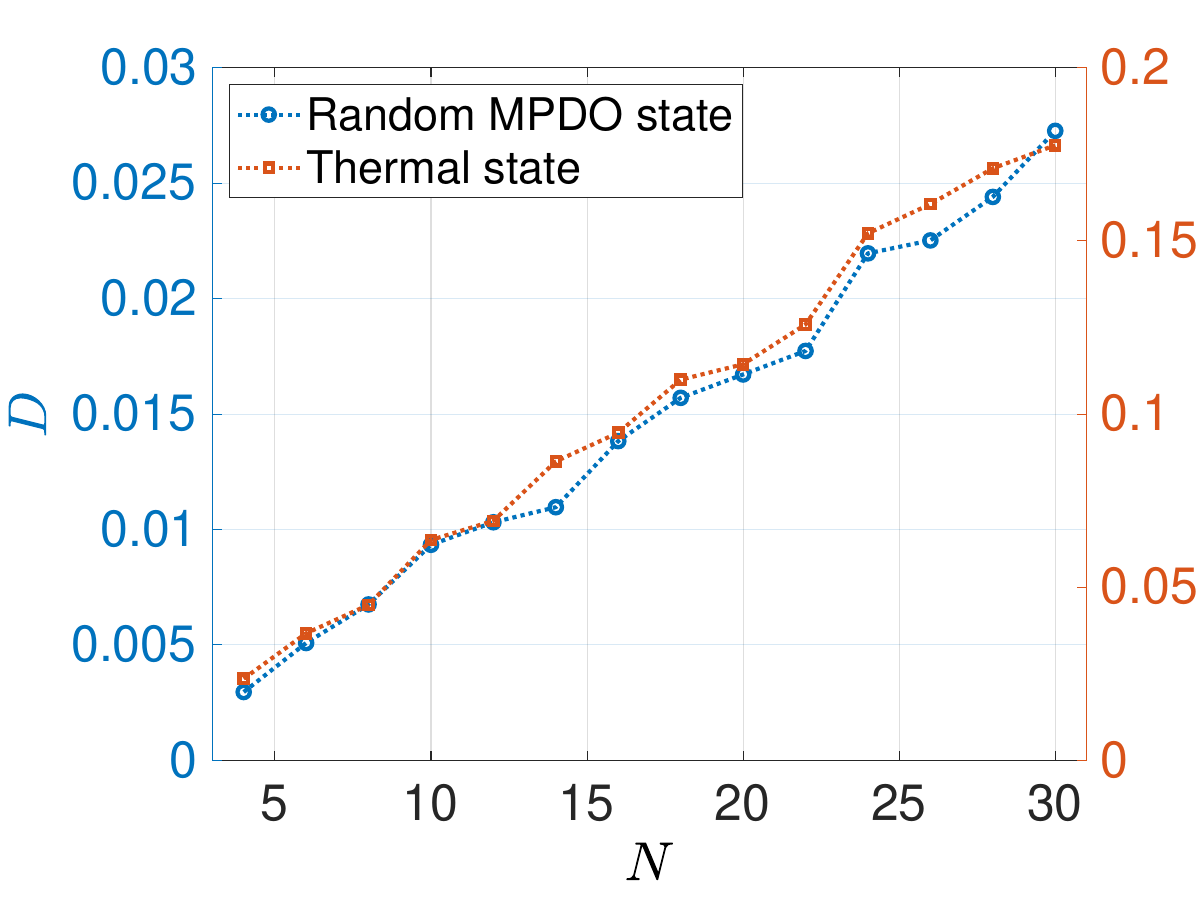}
\caption{The distance measure $\mathcal{D}$ obtained by the MLE algorithm as a function of the number of qubits $N$ for a random MPDO state with $M=10^{4}$ measurement samples for each system size and a thermal state of an 1D quantum Ising model (Eq.\,\eqref{QuantumIsing} with $B=1$ and $T=2$) with $M=1.5\times10^{4}$ measurement samples. The random MPDO state has a bond dimension $2$ and a Kraus dimension $6$, and the model MPDO state shares the same dimensions. The thermal state is represented by an MPO with a maximum matrix dimension of $7$, while the model MPDO used to reconstruct it has a Kraus dimension of 4 and a bond dimension of $3$.\label{fig4}}
\end{figure}

Fig.\,\ref{fig4} indicates that for both the random MPDO states and the thermal states, efficient QST is possible as the distance metric $\mathcal{D}$ scales approximately linearly in $N$. We also point that the performance of our QST protocol that combines the MLE algorithm with the local SIC-POVM compares favorably to existing QST protocols using MPS or MPO \cite{Cramer2010,Baumgratz2013,BaumgratzMLE2013,Wang2020}, as the distance metric we achieved is close to the theoretical limit imposed by the Cramer-Rao bound. Importantly, our QST protocol is fully general and works for any target state, with the number of measurement samples expected to scale only linearly in $N$ for most target states represented by MPS/MPDO with a finite bond dimension.

\section{Conclusion and Outlook}

In this work, we have studied the tomography of a quantum many-body state using a single measurement setting made of local SIC-POVM. This measurement setting can be readily implemented in most current quantum hardware and removes the need of individual qubit controls. We apply the local SIC-POVM measurement to a class of target states that have efficient MPS/MPDO representations and are ubiquitous in one-dimensional physical systems. For most target states in this class, we find efficient QST is possible, with the number of state copies needed to scale linearly with the number of independent parameters in the MPS/MPDO. This conclusion is supported by an efficient calculation of the Cramer-Rao bound and an MLE algorithm using  synthetic measurement samples that saturates the Cramer-Rao bound, hence making the algorithm optimal. We also identify counter examples of efficient QST where the target states belong to a family of GHZ states that are long-range entangled. The number of state copies needed to reconstruct such GHZ states scales exponentially with the system size. Interestingly, one can avoid this exponential scaling by constraining the model and the target state to be both real. This is in contrast to QST protocols based on local reductions, where an ad hoc $N$-qubit string operator needs to be measured to reconstruct the GHZ state \cite{Cramer2010,BaumgratzMLE2013}. Here we do not require the measurement of any non-local operator, since our measurement setting is informationally complete. We have also shown that our MLE algorithm works for both pure target states and mixed target states, making our QST protocol universal.

While these results collectively suggest that the QST protocol proposed in this work can be used towards scalable quantum state tomography for a wide range of physical states in one dimension, a number of open questions remain. First, what is the sufficient and necessary condition for an MPS/MPDO state to be efficiently recoverable under local measurement? We suspect that the answer is closely related to whether the target state is short-range or long-range entangled, as well as to whether we restrict to real wavefunctions. Second, it is not clear whether the MPS/MPDO model provides an unbiased estimation of the target state for a finite sample size. While the Cramer-Rao bound is expected to hold for any well-behaved estimator for a sufficiently large sample size \cite{GS2000,Usami2003}, how to quantify the corrections to the Cramer-Rao bound for a finite sample size remains unclear. Third, can we identify a Cramer-Rao-type bound for a given target state using any local POVM? As we show in the discussion of the generalized GHZ state, for a given target state there may exist an optimal choice of a local POVM to minimize the Cramer-Rao bound. Fourth, we find that the number of samples required to reconstruct a target state accurately is closely related to the number of samples needed to compute the Fisher information matrix $I$ accurately. Could this connection be used to identify states that are not efficiently recoverable by examining the convergence of the Fisher information matrix under an increasing number of Monte Carlo samples? Finally, can the results obtained here be applied to target states with other types of efficient classical representation, such as a higher-dimensional tensor network or an artificial neural network? We believe the study of scalable QST protocols for structured quantum states, and especially the establishment of rigorous bounds on the accuracy of such protocols, is still far from exhausted.

This work is supported by the NSF Grants Nos.\ PHY-2112893, CCF-2106834 and CCF-2241298, as well as the W. M. Keck Foundation. We thank the HPC center at Colorado School of Mines and the Rocky Mountain Advanced Computing Consortium for providing computational resources needed in carrying out this work.

\appendix

\section{Analytical calculation of the Cramer-Rao bound for the GHZ state}\label{GHZCramerRaoSec}

In this appendix, we will calculate the Cramer-Rao bound for the GHZ state analytically. For simplicity, we start with a TI MPS model with the following diagonal matrices:
\begin{equation}
C^{0}=\begin{pmatrix}a_{1} & 0\\
0 & a_{2}
\end{pmatrix}\hspace{1em}C^{1}=\begin{pmatrix}a_{3} & 0\\
0 & a_{4}
\end{pmatrix}
\end{equation}
The use of diagonal matrices here is justified as the inclusion of off-diagonal matrix elements does not change the resulting Cramer-Rao bound at all. The above model state achieves the target state $|\psi_{\text{GHZ}}\rangle=|0\cdots0\rangle+|1\cdots1\rangle$ upon setting $a_{1,4}=1$ and $a_{2,3}=0$.

Since $A^{0}$ and $A^{1}$ commute, we can explicitly calculate the wavefunction of the model MPS state in the computational basis as
\begin{equation}
\psi_{\bm{s}}=\mathrm{Tr}\left(C^{s_{1}}...C^{s_{N}}\right)=a_{1}^{N_{0}}a_{3}^{N-N_{0}}+a_{2}^{N_{0}}a_{4}^{N-N_{0}}
\end{equation}
where $N_{0}$ denotes the number of $0$s in the bit string $\bm{s}$. Let us first assume of $a_{1,2,3,4}$ are real parameters. We can then explicitly evaluate the derivates of $\psi_{\bm{s}}$ over $a_{1,2,3,4}$ as
\begin{align}
\frac{\partial\psi_{\bm{s}}}{\partial a_{1}} & =N\delta_{N_{0},N},\quad\frac{\partial\psi_{\bm{s}}}{\partial a_{2}}=\delta_{N_{0},1}\nonumber \\
\frac{\partial\psi_{\bm{s}}}{\partial a_{3}} & =\delta_{N_{0},N-1},\quad\frac{\partial\psi_{\bm{s}}}{\partial a_{4}}=N\delta_{N_{0},0}\label{eq:dpsida}
\end{align}
where from now on we implicitly assume that all derivatives are further evaluated using the values of $a_{1,2,3,4}$ in the target state. Define the fictitious states $|\Psi_{\alpha}\rangle=\sum_{\bm{s}}\frac{\partial\psi_{\bm{s}}}{\partial a_{\alpha}}|\bm{s}\rangle$ ($\alpha=1,2,3,4$), we find that
\begin{align}
|\Psi_{1}\rangle & =N|0\cdots0\rangle\nonumber \\
|\Psi_{2}\rangle & =|10\cdots0\rangle+|01\cdots0\rangle+|0\cdots01\rangle\nonumber \\
|\Psi_{3}\rangle & =|01\cdots1\rangle+|10\cdots1\rangle+|1\cdots10\rangle\nonumber \\
|\Psi_{4}\rangle & =N|1\cdots1\rangle
\end{align}
which are mutually orthogonal. Next, we notice that Eq.\,\eqref{eq:Kab} can be rewritten as
\begin{align}
K_{\alpha\beta} & =\sum_{\bm{s},\bm{s}^{\prime}}\frac{\partial\left(\psi_{\bm{s}}\psi_{\bm{s}^{\prime}}^{\ast}\right)}{\partial\theta_{\alpha}}\frac{\partial\left(\psi_{\bm{s}}^{\ast}\psi_{\bm{s}^{\prime}}\right)}{\partial\theta_{\beta}}\nonumber \\
 & =2\Re\left(\langle\Psi_{\alpha}|\Psi_{\beta}\rangle\langle\psi|\psi\rangle+\langle\psi|\Psi_{\alpha}\rangle\langle\psi|\Psi_{\beta}\rangle\right)\label{eq:Kabpure}
\end{align}
It's then not hard to find that $K$ is a $4\times4$ matrix with
\begin{align}
K_{11} & =K_{44}=2(2N^{2}+N^{2})=6N^{2}\nonumber \\
K_{22} & =K_{33}=4N\nonumber \\
K_{14} & =K_{41}=2N^{2}
\end{align}
and all other matrix elements of $K$ being zero.

Now we can further assume $a_{1,2,3,4}$ are complex parameters. Note that for any analytic function $f(z)$ with $z=x+iy$, $\frac{df}{dx}=\frac{df}{dz}$ and $\frac{df}{dy}=i\frac{df}{dz}$. As a result, $\frac{\partial\psi_{\bm{s}}}{\partial\Re a_{\alpha}}$ should be equal to Eq.\,\eqref{eq:dpsida} while $\frac{\partial\psi_{\bm{s}}}{\partial\Im a_{\alpha}}$ should be equal to $i$ times Eq.\,\eqref{eq:dpsida}. If $\alpha$ and $\beta$ correspond to a parameter's real part and a parameter's imaginary part respectively, $K_{\alpha\beta}=0$ as we are taking the real part of a purely imaginary number in Eq.\,\eqref{eq:Kabpure}. If $\alpha$ and $\beta$ both correspond to the imaginary parts, the expression of $K_{\alpha\beta}$ is modified to $K_{\alpha\beta}=2\Re\left(\langle\Psi_{\alpha}|\Psi_{\beta}\rangle\langle\psi|\psi\rangle-\langle\psi|\Psi_{\alpha}\rangle\langle\psi|\Psi_{\beta}\rangle\right)$. Treating $K$ now as a $8\times8$ matrices, we have
\begin{align}
K_{55} & =K_{88}=2(2N^{2}-N^{2})=2N^{2}\nonumber \\
K_{66} & =K_{77}=4N\nonumber \\
K_{58} & =K_{85}=-2N^{2}
\end{align}

Noting that $|\psi_{\text{GHZ}}\rangle$ is not properly normalized, we find the normalized $K$ matrix to be:
\begin{equation}
K=\begin{pmatrix}\frac{3}{2}N^{2} & 0 & 0 & \frac{1}{2}N^{2} & 0 & 0 & 0 & 0\\
0 & N & 0 & 0 & 0 & 0 & 0 & 0\\
0 & 0 & N & 0 & 0 & 0 & 0 & 0\\
\frac{1}{2}N^{2} & 0 & 0 & \frac{3}{2}N^{2} & 0 & 0 & 0 & 0\\
0 & 0 & 0 & 0 & \frac{1}{2}N^{2} & 0 & 0 & \frac{-1}{2}N^{2}\\
0 & 0 & 0 & 0 & 0 & N & 0 & 0\\
0 & 0 & 0 & 0 & 0 & 0 & N & 0\\
0 & 0 & 0 & 0 & \frac{-1}{2}N^{2} & 0 & 0 & \frac{1}{2}N^{2}
\end{pmatrix}\label{eq:K_GHZ}
\end{equation}

Next, we will try to calculate the Fisher information matrix for $|\psi_{\text{GHZ}}\rangle$. We can first write $P(\bm{m})$ in Eq.\,\eqref{eq:Pm} as $\Phi(\bm{m})\Phi(\bm{m})^{\ast}$ where
\begin{align}
\Phi(\bm{m}) & =\mathrm{Tr}\left[(F_{0})^{N_{0}}(F_{1})^{N_{1}}(F_{2})^{N_{2}}(F_{3})^{N_{3}}\right]\\
F^{0} & \equiv\frac{1}{\sqrt{2}}C^{0}\nonumber \\
F^{1} & \equiv\frac{1}{\sqrt{6}}C^{0}+\frac{1}{\sqrt{3}}C^{1}\nonumber \\
F^{2} & \equiv\frac{1}{\sqrt{6}}C^{0}+\frac{e^{i2\pi/3}}{\sqrt{3}}C^{1}\nonumber \\
F^{3} & \equiv\frac{1}{\sqrt{6}}C^{0}+\frac{e^{-i2\pi/3}}{\sqrt{3}}C^{1}\nonumber 
\end{align}
Here $N_{q}$ ($q=0,1,2,3$) denotes the number of $q$s in the quart string $\bm{m}=(m_{1},m_{2},\cdots,m_{N})$. We can evaluate $\Phi$ and its derivatives over $a_{1,2,3,4}$ (assuming $a_{1,2,3,4}$ are real for now) in the target state ($a_{1}=a_{4}=1$ and $a_{2}=a_{3}=0$) analytically with the help of Mathematica. The Fisher information matrix can then be evaluated by rewriting Eq.\,\eqref{eq:Iab} as
\begin{equation}
I_{\alpha\beta}=2\Re\sum_{\bm{m}}\left(\frac{\partial\Phi^{\ast}}{\partial\theta_{\alpha}}\frac{\partial\Phi^{\ast}}{\partial\theta_{\beta}}\frac{\Phi}{\Phi^{\ast}}+\frac{\partial\Phi^{\ast}}{\partial\theta_{\alpha}}\frac{\partial\Phi}{\partial\theta_{\beta}}\right) \label{eq:I_sum}
\end{equation}
Importantly, the sum over $\bm{m}$ can be rewritten into the sums over $N_{0,1,2,3}$ that appeared in $\Phi(\bm{m})$ as 
\begin{equation}
\sum_{N_{0}=0}^{N}\sum_{N_{1}=0}^{N-N_{1}}\sum_{N_{2}=0}^{N-N_{0}-N_{1}}\binom{N}{N_{0}}\binom{N-N_{0}}{N_{1}}\binom{N-N_{0}-N_{1}}{N_{2}}
\end{equation}
where we shall set $N_{3}=N-N_{0}-N_{1}-N_{2}$ in the summand of Eq.\,\eqref{eq:I_sum}. This allows us to perform the summation over $\text{\ensuremath{\bm{m}}}$ efficiently.

Now assuming $a_{\alpha}$ ($\alpha=1,2,3,4$) is complex, similar to how we deal with the $K$ matrix, here we find that $\frac{\partial\Phi}{\partial\Re a_{\alpha}}$ is the same as $\frac{\partial\Phi}{\partial a_{\alpha}}$ with $a_{\alpha}$ assumed real, while $\frac{\partial\Phi}{\partial\Im a_{\alpha}} =i\frac{\partial\Phi}{\partial\Re a_{\alpha}}$. With the help of these analytical derivations, we can compute the Fisher information matrix numerically. After a proper normalization for the target state $|\psi_{\text{GHZ}}\rangle$, we obtain

\begin{equation}
I=\begin{pmatrix}\gamma N^{2} & 0 & 0 & \delta N^{2} & 0 & 0 & 0 & 0\\
0 & N & 0 & 0 & 0 & 0 & 0 & 0\\
0 & 0 & N & 0 & 0 & 0 & 0 & 0\\
\delta N^{2} & 0 & 0 & \gamma N^{2} & 0 & 0 & 0 & 0\\
0 & 0 & 0 & 0 & \delta N^{2} & 0 & 0 & -\delta N^{2}\\
0 & 0 & 0 & 0 & 0 & N & 0 & 0\\
0 & 0 & 0 & 0 & 0 & 0 & N & 0\\
0 & 0 & 0 & 0 & -\delta N^{2} & 0 & 0 & \delta N^{2}
\end{pmatrix}\label{eq:I_GHZ}
\end{equation}
where $\gamma\approx2$ and $\delta\approx1/2^{N}$ are asymptotically exact in the large $N$ limit. If the parameters in the MPS model are assumed real, we can just restrict the calculated $K$ and $I$ matrices {[}Eq.\,\eqref{eq:K_GHZ} and Eq.\,\eqref{eq:I_GHZ}{]} to the first four rows and columns, which correspond to the real part of the $a_{1,2,3,4}$.

Next, we generalize the above calculations to a non-TI MPS model, where
\begin{equation}
C_{i}^{0}=\begin{pmatrix}a_{i,0} & 0\\
0 & b_{i,0}
\end{pmatrix}\hspace{1em}C_{i}^{1}=\begin{pmatrix}a_{i,1} & 0\\
0 & b_{i,1}
\end{pmatrix}
\end{equation}
which reduces to the target state upon setting $a_{i,0}=b_{i,1}=1$ and $b_{i,0}=a_{i,1}=0$ for all $i$. It's easy to see that 
\begin{equation}
\psi_{\bm{s}}=\mathrm{Tr}\left(A_{1}^{s_{1}}...A_{N}^{s_{N}}\right)=\prod_{i=1}^{N}a_{i,s_{i}}+\prod_{i=1}^{N}b_{i,s_{i}}
\end{equation}
Following similar derivations as the TI case above, we can obtain the normalized $K$ matrix in a block form as
\begin{equation}
K=\begin{pmatrix}\frac{3}{2}\bm{1} & \bm{0} & \bm{0} & \frac{1}{2}\bm{1}\\
\bm{0} & \mathbb{I} & \bm{0} & \bm{0}\\
\bm{0} & \bm{0} & \mathbb{I} & \bm{0}\\
\frac{1}{2}\bm{1} & \bm{0} & \bm{0} & \frac{3}{2}\bm{1}
\end{pmatrix}\label{eq:K_TIGHZ}
\end{equation}
where $\bm{1}$ is an $N\times N$ matrix of all 1s, $\bm{0}$ an $N\times N$ matrix of all 0s, and $\mathbb{I}$ an $N\times N$ identity matrix. Similarly, we can obtain the Fisher information matrix in a block form as:

\begin{equation}
I=\begin{pmatrix}\gamma\bm{1} & \bm{0} & \bm{0} & \delta\bm{1} & \bm{0} & \bm{0} & \bm{0} & \bm{0}\\
\bm{0} & \mathbb{I} & \delta\mathbb{I} & \bm{0} & \bm{0} & \bm{0} & \bm{0} & \bm{0}\\
\bm{0} & \delta\mathbb{I} & \mathbb{\mathbb{I}} & \bm{0} & \bm{0} & \bm{0} & \bm{0} & \bm{0}\\
\delta\bm{1} & \bm{0} & \bm{0} & \gamma\bm{1} & \bm{0} & \bm{0} & \bm{0} & \bm{0}\\
\bm{0} & \bm{0} & \bm{0} & \bm{0} & \delta\bm{1} & \bm{0} & \bm{0} & -\delta\bm{1}\\
\bm{0} & \bm{0} & \bm{0} & \bm{0} & \bm{0} & \mathbb{I} & \bm{0} & \bm{0}\\
\bm{0} & \bm{0} & \bm{0} & \bm{0} & \bm{0} & \bm{0} & \mathbb{I} & \bm{0}\\
\bm{0} & \bm{0} & \bm{0} & \bm{0} & -\delta\bm{1} & \bm{0} & \bm{0} & \delta\bm{1}
\end{pmatrix}\label{eq:I_TIGHZ}
\end{equation}
Here $\gamma$ and $\delta$ take exactly the same $N$-dependent values as the TI case. If the parameters in the model are assumed real, we can again simply restrict the $K$ and $I$ matrices obtained to the first four rows and columns in the above block forms.

With these analytical derivations, one can now efficiently calculate $\text{Tr}(KI^{-1})$ for both the TI and the non-TI MPS model and with either real or complex parameters. A similar but more involved calculation can be done for the generalized GHZ state in Eq.\,\eqref{eq:gGHZ}. The results of these calculations are shown in Fig.\,\ref{fig2} of the main text.

\bibliographystyle{apsrev4-1}
\bibliography{refs,library}

\end{document}